\documentclass[twocolumn]{aastex63}
\usepackage{amsmath}
\usepackage{hyperref}
\usepackage[T1]{fontenc}
\usepackage{xcolor}
\usepackage{soul}
\usepackage{subfigure}
\usepackage{graphicx}
\usepackage{todonotes}
\usepackage[version=4]{mhchem}
\usepackage{isotope}
\usepackage{multirow}
\usepackage{tablefootnote}
\newcommand{\rp}{\emph{r}-process}
\newcommand{\ye}{$\rm{Y_e}$}
\newcommand{\iso}[2]{\ensuremath{^{#2}}#1}

\defcitealias{Zhu2021}{Z21}
\defcitealias{Barnes2021}{B21}

\newcommand{\mzero}{\textsf{nuclearmodel.ye}}
\newcommand{\mone}{\textsf{sly4.18}}
\newcommand{\mtwo}{\textsf{etfsi.18}}
\newcommand{\mthree}{\textsf{hfb22.02}}
\newcommand{\mfour}{\textsf{hfb27.02}}
\newcommand{\mfive}{\textsf{hfb27.18}}

\begin{document}
\title{The Influence of $\beta$-Decay Rates on r-Process Observables}

\author[0000-0003-0031-1397]{Kelsey A. Lund}
\affiliation{Department of Physics, North Carolina State University, Raleigh, NC 27695 USA}
\affiliation{Theoretical Division, Los Alamos National Laboratory, Los Alamos, NM 87544 USA}
\affiliation{Joint Institute for Nuclear Astrophysics—Center for the Evolution of the Elements, USA}

\author[0000-0002-2748-6640]{J. Engel}
\affiliation{Department of Physics and Astronomy, University of North Carolina, Chapel Hill, NC 27599 USA}

\author[0000-0001-6811-6657]{G.C. McLaughlin}
\affiliation{Department of Physics, North Carolina State University, Raleigh, NC 27695 USA}
\affiliation{Joint Institute for Nuclear Astrophysics—Center for the Evolution of the Elements, USA}

\author[0000-0002-9950-9688]{M.R. Mumpower}
\affiliation{Theoretical Division, Los Alamos National Laboratory, Los Alamos, NM 87544 USA}
\affiliation{Center for Theoretical Astrophysics, Los Alamos National Laboratory, Los Alamos, NM 87544 USA}
\affiliation{Joint Institute for Nuclear Astrophysics—Center for the Evolution of the Elements, USA}

\author[0000-0001-8008-2314]{E.M. Ney}
\affiliation{Department of Physics and Astronomy, University of North Carolina, Chapel Hill, NC 27599 USA}

\author[0000-0002-4729-8823]{R. Surman}
\affiliation{University of Notre Dame, Notre Dame, IN 46556 USA}
\affiliation{Joint Institute for Nuclear Astrophysics—Center for the Evolution of the Elements, USA}

\begin{abstract}
The rapid neutron capture process (r-process) is one of the main mechanisms whereby elements heavier than iron are synthesized, and is entirely responsible for the natural production of the actinides. Kilonova emissions are modeled as being largely powered by the radioactive decay of species synthesized via the \rp. Given that the \rp{} occurs far from nuclear stability, unmeasured beta decay rates play an essential role in setting the time scale for the \rp. In an effort to better understand the sensitivity of kilonova modeling to different theoretical global beta-decay descriptions, we incorporate these into nucleosynthesis calculations. We compare the results of these calculations and highlight differences in kilonova nuclear energy generation and light curve predictions, as well as final abundances and their implications for nuclear cosmochronometry. We investigate scenarios where differences in beta decay rates are responsible for increased nuclear heating on time scales of days that propagates into a significantly increased average bolometric luminosity between 1-10 days post-merger. We identify key nuclei, both measured and unmeasured, whose decay rates are directly impact nuclear heating generation on timescales responsible for light curve evolution. We also find that uncertainties in beta decay rates significantly impact ages estimates from cosmochronometry.
\end{abstract}

\keywords{r-process,beta decay,kilonova,neutron stars}

\section{Introduction}\label{introduction}

Since its emergence in the 1950s, one of the biggest questions in the field of nuclear astrophysics remains the main production site of some of the heaviest elements, synthesized via the rapid neutron capture process (\rp) \citep{Burbidge1957,Cameron1957}. It is currently hypothesized that this process, thought to be responsible for roughly half the material heavier than iron, as well as the only process for producing the actinides, occurs to some extent in the neutron-rich outflows of neutron star mergers (NSM). In addition to the many theoretical advances made in the last few decades, the recent electromagnetic observations accompanying the gravitational wave event GW170817 \citep{Abbott2017,Abbott2017a,Diaz2017} lend the most support to the long-standing idea of a kilonova (KN) explosive transient powered by the radioactive decay of freshly synthesized \rp{} material \citep{Lattimer1974,Lattimer1976,Li1998,Metzger2010,Roberts_2011,Barnes2013,grossman_2014,wollaeger_2018,fontes_2020}.

The luminosity and morphology of the light curve associated with AT2017gfo \citep{Chornock2017,Cowperthwaite2017,Nicholl2017,Perego_2017} offer unique insight into the physics of these extreme environments. The bright but rapidly decaying component of the light curve observed at shorter wavelengths indicates at least some portion of ejecta material with little to no lanthanide or actinide abundances \citep{Metzger2010,Roberts_2011,Evans2017,Miller2019}. On the other hand, a dimmer "red" signal that dominates on timescales of days (when the "blue" signal has faded away) indicates that at least some portion of the ejected material is composed of high-opacity lanthanides and possibly actinides \citep{Barnes2013,Tanaka2013,Kasen2017}. These combined observations suggest distinct nucleosynthesis sites within the merger ejecta, each of which might be capable of producing a robust \rp{} pattern.

Despite the wealth of information provided by the data from GW170817, the larger endeavour of modeling KN signals for the purpose of reliably interpreting future signals remains subject to many unknown quantities and large uncertainties. While it is generally accepted that NSMs are a site for \rp{} production, it remains unclear whether these sites alone are capable of producing the entire observed \rp{} pattern. Studies of material ejected on dynamical timescales, either via tidal forces or compression between the coalescing neutron stars, predict material that is neutron-rich enough to produce out to the second and third \rp{} peaks \citep{Rosswog_1999,Goriely2011,Korobkin2012,Bauswein2013,Wanajo2014,Sekiguchi2015,Radice2018}.

While there is general agreement on the robustness of \rp{} production in dynamically ejected channels, there is more uncertainty regarding the extent to which \rp{} production occurs in in late-time accretion disk outflows driven by viscous heating. Some part of this uncertainty comes from the central remnant scenario \citep{shibata_2005,agathos_2020,Nedora_2021,kashyap_2022}. When prompt collapse does not occur, neutrino interactions are capable of driving the \ye{} up enough to stifle \rp{} production. Neutrino oscillations can also play a large part in determining the extent of this effect, as only electron neutrinos act to reduce the neutron-richness of the ejecta \citep{Malkus2012,Siegel2017,Li2021}.

Simulating \rp{} nucleosynthesis is also subject to large uncertainties due to its trajectory far from nuclear stability where many important quantities remain unmeasured \citep{Mendoza-Temis2015,eichler_2015,Mumpower2016a,nikas_2020}. Detailed calculations incorporate nuclear heating contributions from multiple decay modes, and these impact the energy released, the thermalization efficiency with which the decay products deposit energy into the system, and the composition of material that is ultimately synthesized \citep{Beun_2008,Barnes2016,mumpower2018,Even2019,sprouse_2020}. 

The incorporation of theoretical spontaneous fission and alpha decay rates have been found to largely impact the uncertainty in the nuclear heating on time scales relevant for the evolution of the light curve, in some cases adding up to an order of magnitude to the uncertainty (\citet{Vassh2019}; \citet{Giuliani2020}; \citetalias{Zhu2021}; \citetalias{Barnes2021}). These combine with additional uncertainties including the nuclear equation of state \citep{Oechslin2007,Bauswein2013,Hotokezaka2013,Sekiguchi2015,Lehner2016}, the nature of neutrino oscillations and neutrino transport in the ejecta \citep{Miller2019,Kullmann2021}, as well as atomic line energy calculations for high-opacity lanthanides and actinides.

One particularly important data set for an \rp{} nucleosynthesis calculation is a description of the beta decay rates involved \citep{Moller2003,caballero_2014,Marketin2016,Shafer2016,Ney2020,Robin_2022,kullmann_2022}. At early times, the extent of r-process production is sensitive to the beta decay rates of the nuclei involved, as these determine the relative abundances of connected isotopic chains during $(n,\gamma)-(\gamma,n)$ equilibrium and compete directly with neutron capture when equilibrium fails. At later times, and once a population of high mass number species is synthesized, beta decay rates play a further role in determining the time scale of beta decay chains which are important for heating as well as for populating species which contribute significantly to spontaneous fission and alpha decay heating. Additionally, theoretical beta decay rates can compete with theoretical alpha decay and spontaneous fission branching ratios which are crucial for determining the shape and magnitude of the light curve \citetalias{Zhu2021}. 

We aim to incorporate different global beta decay descriptions into nucleosynthesis calculations and compare their impact with those of other astrophysical and nuclear sources of uncertainty on nuclear energy generation, light curve evolution, and predictions relevant to nuclear cosmochronometry. In section \ref{sec:method}, we describe the methods we use in generating and compiling nuclear data as well as the computational methods we use for calculating relevant quantities. In section \ref{sec:effheat}, we show the results of our calculations of nuclear energy generation. We also show the impact these uncertainties in nuclear effective heating can have on a bolometric light curve. We conclude the presentation of our results with section \ref{sec:cosmoresult}, which makes age estimates for several \rp-enhanced metal poor stars. Finally, we provide some concluding remarks in section \ref{sec:conclusion}.

\section{Method}\label{sec:method}

We seek to quantify the leverage of beta decay rates on key aspects of kilonova modeling when compared with other sources of nuclear and astrophysical uncertainty. The evolution of the nucleosynthetic abundances throughout the \rp{} determines the energy output in the form of nuclear heating. The thermalization profile of this released energy will determine how it is transported away from the system, which in turn affects the shape and magnitude of the observable light curve. The final abundance that is produced in a given merger even can then be used in a stellar dating technique if it is interpreted as being the sole source of a star's \rp{} material. In this section, we describe our data set and the methods used in the calculation of these quantities.

\subsection{Model Set and Nucleosynthesis}\label{sec:models}
We use the Portable Routines for Integrated nucleoSynthesis Modeling (PRISM) to perform nucleosynthesis calculations using a suite of prepared input files describing astrophysical conditions and nuclear properties for a wide range of nuclei. The extent of \rp{} production is sensitive to the beta decay rates of the nuclei involved, as these compete with neutron capture rates. Currently, many methods exist to compute beta decay rates, but few are applied to large sections of the chart of the nuclides. In order to investigate the extent of the impact of different sets of beta decay rates in our calculations, we construct separate beta decay and coupled beta decay/beta-delayed fission reaction data sets consistent with three different beta decay calculations.

\begin{figure}
    \includegraphics[scale=0.45]{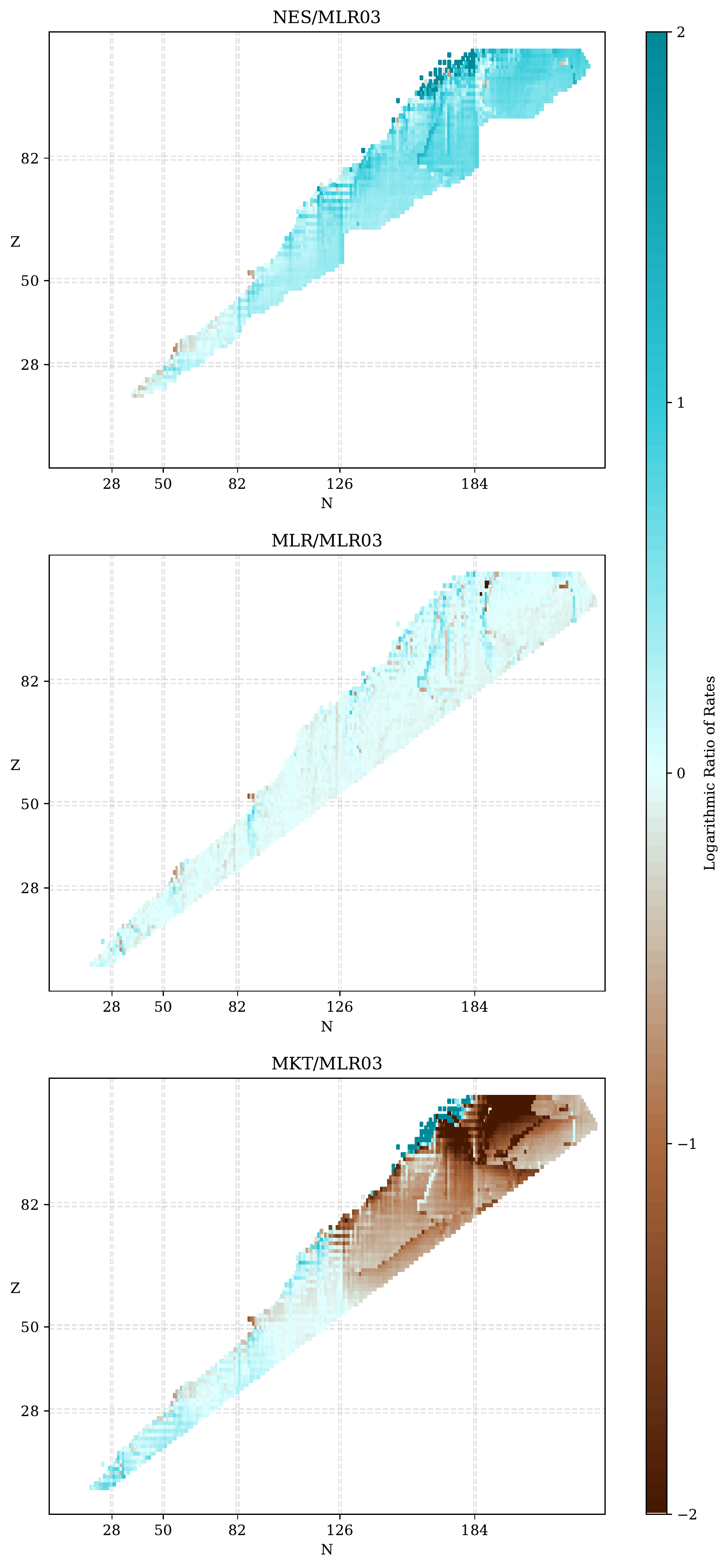}
    \caption{Logarithmic ratios of NES (left), MLR (center), and MKT (right) beta decay rates with respect to MLR03 beta decay rates. The color bar is adjusted to values of -2,2 to reveal trends more clearly. The area covered by the Nubase 2016 data set \citep{Audi2017} is removed.}
    \label{fig:rateratios} 
\end{figure}

The description contained in the work of \citet{Ney2020} (calculations using these rates will hereafter be referred to as NES) uses the finite amplitude method with Skyrme density functionals to compute the beta decay half-lives for neutron rich species. The work of \citet{Marketin2016} (hereafter MKT) uses a covariant density functional theory approach with Gogny interactions to do the same. \citet{Moller2019} (hereafter MLR) uses a finite range droplet model in a quasiparticle random phase approximation to obtain $\beta$-strength functions for neutron-rich species. We use the three sets of beta decay rates described and compute beta delayed neutron emission and beta-delayed fission probabilities and daughter product distributions using \citet{Mumpower2016}. We show the base-ten logarithm of the ratio of all three sets of beta decay rates with respect to those of \citet{Moller2003} (MLR03) in Figure \ref{fig:rateratios}. 

In addition to beta decays, the nuclei involved in the \rp{} are also subject to other reactions and decays. The energy associated with these decays is important for the nuclear heating, and plays a large part in determining the shape and magnitude of the light curve. We calculate reaction and decay rates, as well as Q-values, consistent with the eight nuclear models listed in Table \ref{tab:masses}. We incorporate theoretical alpha decay rates obtained using a Viola-Seaborg relation. We use neutron capture and neutron-induced fission rates calculated using the statistical Hauser-Feshbach code, CoH \citep{Kawano2016}.

For spontaneous fission rates, we use the barrier-height-dependent prescription from \citet{karpov2012} and \citet{Zagrebaev2011}. We adopt mass models with appropriate fission barrier height descriptions: ETFSI \citep{Aboussir1995} with ETFSI, TF \citep{Myers1999} with TF, HFB14 \citep{Goriely2009} with HFB22 and HFB27, and FRLDM \citep{Moller2015} for all others. We consider two possible fission fragment distributions; the first is a symmetric split, where the daughter products each equal one-half of the parent nucleus, while the second is the double Gaussian distribution described by \citet{Kodama1975} (hereafter K\&T). We make an exception for the fission fragment distribution of the spontaneous fission of \iso{Cf}{254}, where we use the more detailed calculation from \citet{Zhu2018}.

Finally, where experimental or evaluated data is available, we overwrite theory rates with data from Nubase \citep{Audi2017}, and calculate Q-values using experimentally determined masses from AME2016 \citep{Wang2017}.

\begin{deluxetable}{l r}
\tablewidth{\textwidth}
\tablecaption{\label{tab:masses}Nuclear mass models (and associated references) used in nucleosynthesis calculations.}
\tablehead{\colhead{Abbreviation} & \colhead{References}}
\startdata
    DZ33 & \citet{Duflo1995}\\
    ETFSI & \citet{Aboussir1995,Mamdouh2001},\\
    FRDM2012 & \citet{Moller2016}\\
    HFB22,HFB27 & \citet{Goriely2009,Goriely2013a}\\
    SLY4 & \citet{Chabanat1998,Moller2015}\\
    TF & \citet{Myers1996,Myers1999}\\
    UNEDF1 & \citet{Kortelainen2012,Moller2015}\\
    WS3 & \citet{Liu2011,Moller2015}\\
\enddata
\end{deluxetable}

For the thermodynamic evolution of the ejecta, we use a parameterized wind model with an initial entropy per baryon of $s/k=40$ and an expansion timescale of 20 ms. The network begins in nuclear statistical equilibrium with initial seed nuclei determined using the SFHo equation of state \citep{Steiner2013}. We use the initial electron fraction, \ye, as a proxy of variation in the astrophysical conditions of the ejecta in order to compare with uncertainties from the previously described variations from theoretical nuclear models. We use single-\ye{} trajectories with initial values of 0.02, 0.18, and 0.21 for the full suite of theoretical nuclear inputs. Based on the work in \citet{Zhu2021}, we consider that these can be taken to represent varying degrees of contribution from fission to the total heating, with $\rm{Y_{e,i}}=0.21$ yielding the smallest contribution.

\begin{figure}
    \centering
    \includegraphics[scale=0.36]{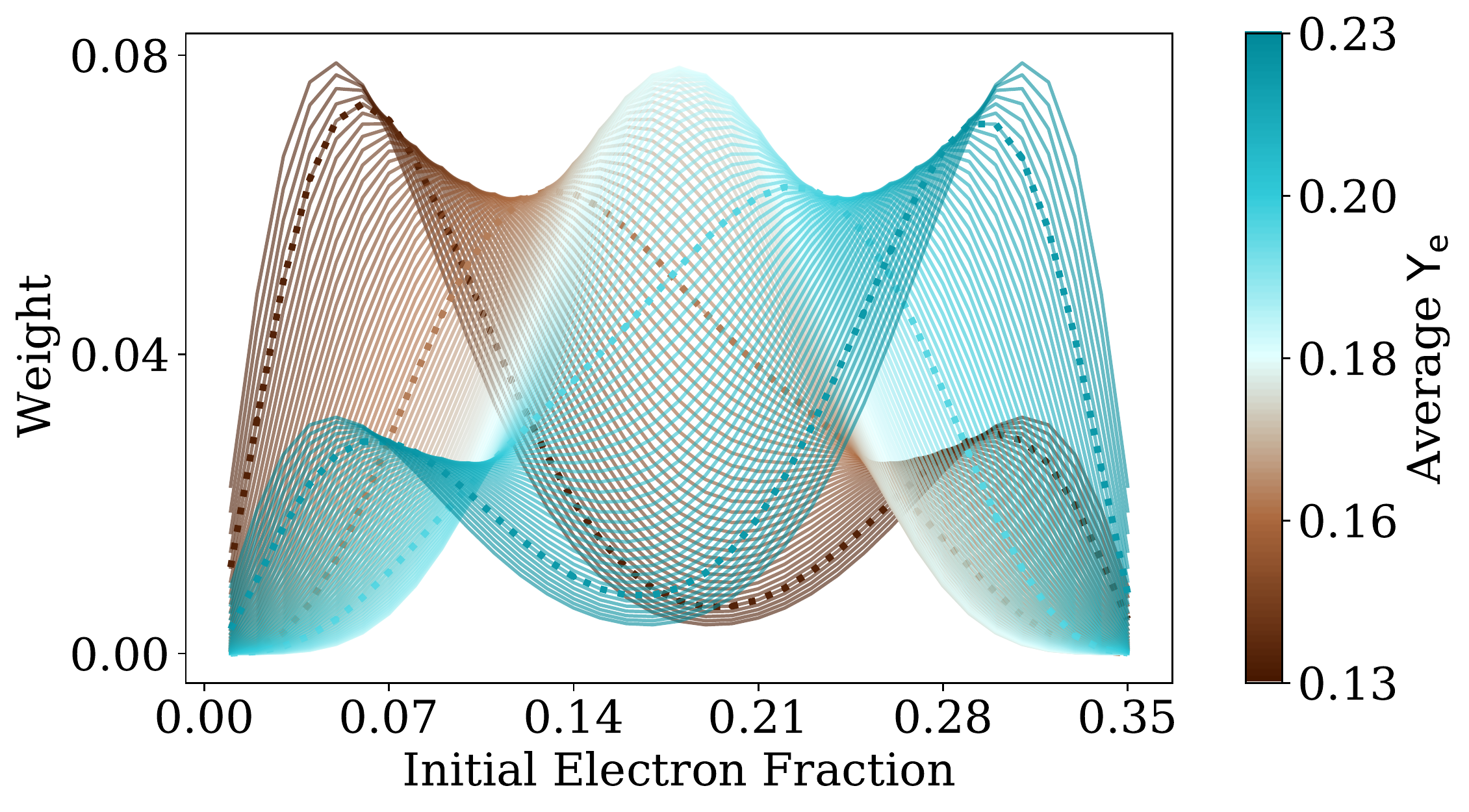} 
    \caption{Distribution of combined single-\ye{} trajectories. The color of each line indicates the average \ye{} of the "combined trajectory". A selection of combinations are shown as a guide to highlight the double-peak structures within the distribution. This color scheme is used throughout this work to refer to results for each combined trajectory, with the bluer combinations having less neutron-rich (low \ye) material and the more brown combinations having more neutron-rich material.}
    \label{fig:ye_dist}
\end{figure}

\begin{figure*}
    \centering
    \includegraphics[scale=0.36]{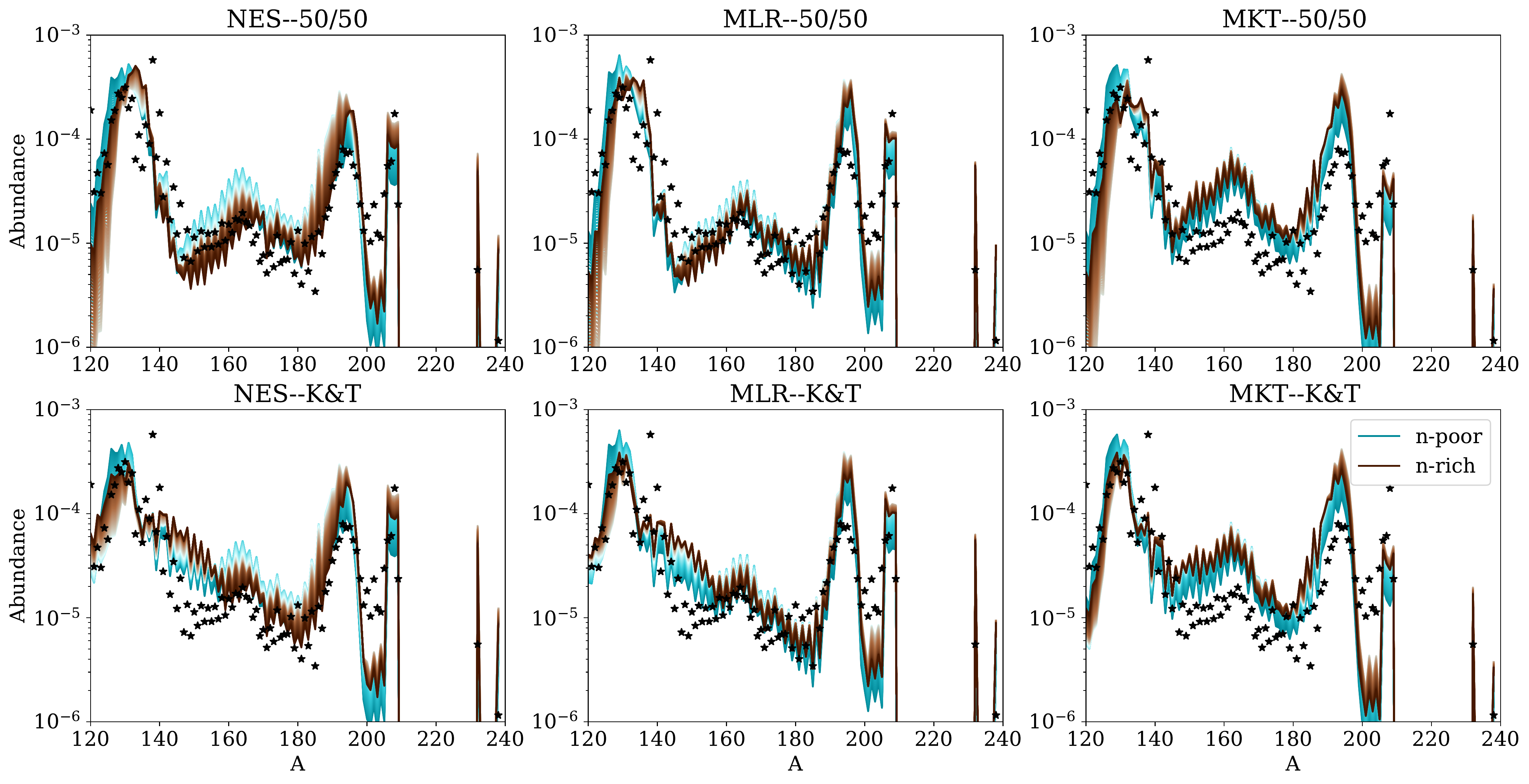} 
    \caption{Abundance patterns for combined trajectories displayed in Figure \ref{fig:ye_dist}. The coloring convention follows that of Figure \ref{fig:ye_dist}, with brown representing more neutron-rich combinations, and brown more neutron-poor. Solar abundances \citep{asplund_2009} are plotted as black stars.}
    \label{fig:ab_yedist}
\end{figure*}

We also consider a set of trajectories to more closely model an ejecta with non-uniform composition. To do this, we perform nucleosynthesis calculations using the FRDM2012 subset of nuclear inputs on single-\ye{} trajectories ranging from 0.01 to 0.35 in increments of 0.01. We map these onto an analytic probability distribution in order to sample a range of distributions. We show these mappings in Figure \ref{fig:ye_dist}. Previous work found that solar-like abundances can be obtained by combining individual trajectories with both high initial \ye{} and well as low initial \ye{} (\citetalias{Zhu2021}. Our goal is not necessarily to obtain a solar-like final abundance pattern (we direct the reader to, for example, \citet{ristic_2022}, which addresses \rp{} universality), since we do not know if kilonova events produce a solar pattern, but rather to sample a variety of combined trajectories containing material with a variety of neutron richness. This is consistent with most modern kilonova models predict multiple ejecta components (see for example \citet{Radice2018}; \citet{Miller2019}; \citet{Nedora_2021}; \citet{stewart_2022}). We utilize these combinations of simulations for effective heating, as well as light curve calculations out to 50 days, and use the final (1 Gyr) abundances for cosmochronometry calculations (see section \ref{sec:cosmo}).  We show these resulting abundance patterns in Figure \ref{fig:ab_yedist}.

\subsection{Nuclear Heating}\label{sec:methodNucHeat}
The evolution of the nuclear heating is given by coupling the energy from the radioactive decays with the efficiency with which their products thermalize. We therefore take the outputs from the nucleosynthesis calculations described and calculate the effective heating rate by combining the Q-value from the radioactive decays with the thermalization efficiency. We follow the method in \citet{Kasen2019} and \citetalias{Zhu2021}, with the total effective heating rate given by
\begin{equation}
    \dot{Q}(t) = \sum_{i} \dot{q}_i(t) f_i (M_{ej},v_{ej},t) M_{ej},
\end{equation}
where the sum is over all reactions and decays. The thermalization, $f_i$, is calculated as in \citet{Kasen2019} and is dependent on the ejecta mass and velocity. We use a total ejecta mass of $0.05 M_{\odot}$ and an ejecta velocity of $0.15c$.

\subsection{Semi-Analytic Light Curve Model}\label{sec:methodLC}
We construct a semi-analytic light curve model following \citetalias{Zhu2021} and \citet{Metzger2020}. We include some of the relevant details of the calculation here for convenience. We divide the ejecta into 100 layers with $0.1c<v<0.4c$, with a velocity-dependent mass distribution and density profile given by
\begin{align}
    M_v &= M_{ej}\left (\frac{v_o}{v}\right )^3\\
    \rho(v,t) &= \frac{3M_v}{4\pi v^3t^3}.
    \label{eq:shellmass}
\end{align}
The evolution of each shell is calculated independently using a forward-Euler scheme from 0.01 days to 50 days post-merger, obeying 
\begin{equation}
    \frac{dE_v}{dt}=\frac{M_v}{M_{ej}}\dot{Q}(t,v)-\frac{E_v}{t}-L_v.
    \label{eq:evol}
\end{equation}
The last term,
\begin{equation}
    L_v=\frac{E_v}{t_{d,v}+t_{lc}},
    \label{eq:lumv}
\end{equation}
is the luminosity of a shell and is dependent on the diffusion timescale, \textit{$t_{d,v}$}, and the light-crossing time, \textit{$t_{lc}$} of that shell:
\begin{align}
    t_{\rm d,v} &= \frac{M_{ext}\kappa}{4\pi vtc}, \: \text{ and} \\
    t_{\rm lc} &= \frac{vt}{c}.
\end{align}
Here, $M_{ext}$ is the mass exterior to the shell with velocity v. $\kappa$ is the opacity of the layer, which is calculated as a function of the temperature, $T_v$, 
\begin{equation}
    T_v = \left (\frac{E_v \rho(v,t)}{a M_{norm}}\right )^{\frac{1}{4}}\\
\end{equation}
of the layer, as:
\begin{equation}
\kappa =
\begin{cases}
\kappa_{\rm max} \left(\frac{T}{4000 \; \rm{K}}\right)^{5.5}, \: &\text{ $T < 4000$ K } \\
\kappa_{\rm max} \: &\text{ otherwise }
\end{cases}
\end{equation}

The value of $\kappa_{\rm max}$ depends on the composition of the ejecta. Specifically, the presence of lanthanides and actinides contribute large opacities which are important for the "red" component of the light curve that is relevant on timescales of days (and the focus of this work). Given the temperature-dependent treatment we have selected (as opposed to a "grey" opacity), we adopt a value of $\kappa_{\rm max}$ of $100\, \rm cm^2g^{-1}$ for all simulations, as this represents the scale of the maximum opacity from more detailed calculations of low-\ye{}, lanthanide-bearing ejecta on timescales of a few days \citep{Kasen2013,Tanaka2020}.

\subsection{Nuclear Cosmochronometry}\label{sec:cosmo}

\begin{deluxetable*}{l c c c r}\label{tab:stars}
\tablewidth{0pt}
\tablecaption{ Names of select r-ii stars with their observed abundances. These stars are sorted by increasing actinide enhancement.}
\tablehead{\colhead{Star Name} & \colhead{$\rm{log_\epsilon(Eu)}$} & \colhead{$\rm{log_\epsilon(Th)}$} & \colhead{$\rm{log_\epsilon(U)}$} & \colhead{Reference(s)}}
\startdata
    CS22892-052 & -0.95 & -1.57 & -2.3 & \citet{Sneden2003}\\
    HE1523-0901 & -0.62 & -1.2 & -2.06 & \citet{Frebel2007}\\
    CS29497-004 & -0.66 & -1.16 & -2.20 & \cite{Hill2017}\\
    CS31082-001 & -0.72 & -0.98 & -1.92 & \citet{SiquieraMello2013}\\
    & & & & \citet{Hill2002}\\
    J2038-0023 & -0.75 & -1.24 & -2.14 & \citet{Placco2017}\\
    J0954+5246 & -1.19 & -1.31 & -2.13 & \citet{Holmbeck2018obs}\\
\enddata
\end{deluxetable*}

The material produced in the NSM can act as an enrichment source for a nearby stellar environment, i.e. the final abundance of an \rp{}-producing event can be taken as the initial abundance of a star. These can then be compared with spectral observations of stars, and a time can be obtained by comparing the measured decay timescales and abundances of radioactive species. To this end, we make a selection of six \rp{}-enhanced ($[\rm{Eu}/\rm{Fe}]>+1.0$), metal-poor ($[\rm{Fe}/\rm{H}]<-2$) stars of varying actinide richness, listed in Table \ref{tab:stars}. 

From these observations, we use the measurements of europium (Z=63), thorium (Z=90), and uranium (Z=92). We take the initial abundances to be those produced in a NSM and remaining 1 Gyr post-merger. Running the calculations out this long allows us to only rely on the theoretical abundances of the long-lived isotopes of thorium (\iso{Th}{232}) and uranium (\iso{U}{238}),which have half-lives of 14 Gyr and 4.47 Gyr, respectively. We compare these with the final abundances of the two most stable isotopes of europium (\iso{Eu}{151} and \iso{Eu}{153}).

If the NSM is taken to be the sole source of \rp{} enrichment, and is interpreted as occurring at t=0, then the observed spectra can be interpreted as being taken at time t given by the following relations: 
\begin{align}
    t &= 46.67\ \rm{Gyr}\ [log_\epsilon(Th/Eu)_0-log_\epsilon(Th/Eu)_{\rm{obs}}]\label{eq:ThEu}\\
    t &= 14.84\ \rm{Gyr}\ [log_\epsilon(U/Eu)_0-log_\epsilon(U/Eu)_{\rm{obs}}]\label{eq:UEu}\\
    t &= 21.80\ \rm{Gyr}\ [log_\epsilon(U/Th)_0-log_\epsilon(U/Th)_{\rm{obs}}]\label{eq:UTh}.
\end{align}
These compare the initial abundances, $\rm{log_\epsilon (X/Y)_0}$, to abundances, $\rm{log_\epsilon (X/Y)_{\rm{obs}}}$. This approach, while useful, has a tendency to yield inconsistent results, especially when applied to actinide-boost stars, which are overabundant in thorium and uranium (\citet{Holmbeck_2019} classifies these as having $\rm log_\epsilon(Th/Dy)>-0.90$).

We explore the impact of changing the description of beta decay rates on the final abundance pattern where relevant for cosmochronometry calculations using the stars listed in Table \ref{tab:stars}. The abundance patterns we use to perform these calculations are constructed from the subset of individual trajectories using the FRDM2012 mass model, with initial electron fractions ranging from 0.01 to 0.35, as described in section \ref{sec:models}.

\section{Effective Heating and Light Curve}\label{sec:effheat}

We demonstrate the influence of beta decay rates on the effective heating and the light curve by first considering single trajectory models which have a single value of the initial electron fraction. We then turn to multi-trajectory models which account for ejecta which has a weighted range of initial neutron richness, as illustrated in Figure \ref{fig:ye_dist}. 
In all cases, we take the total heating to be the summed contribution of the effective heating (as described in Section \ref{sec:methodNucHeat}) from beta decay, spontaneous fission, and alpha decay reactions. 

\subsection{Single Trajectory}\label{sec:single}
We begin with three different single trajectory models which are chosen to access different physics. The \ye$=0.02$ case is chosen to probe very neutron rich ejecta that experiences fission cycling, where the daughter products of the first nuclei which fission capture enough neutrons to make it back to very heavy nuclei that will fission again. The \ye$=0.18$ case is chosen because a significant number of nuclei fission, but there is  limited cycling, since at this \ye{}, the number of neutrons is not enough to allow nuclei to fission twice. Finally, the \ye$=0.21$ case was chosen because material with this neutron richness makes a full r-process but does not have enough neutrons for much fission to occur.

\begin{figure*}[ht!]
    \includegraphics[width=\textwidth]{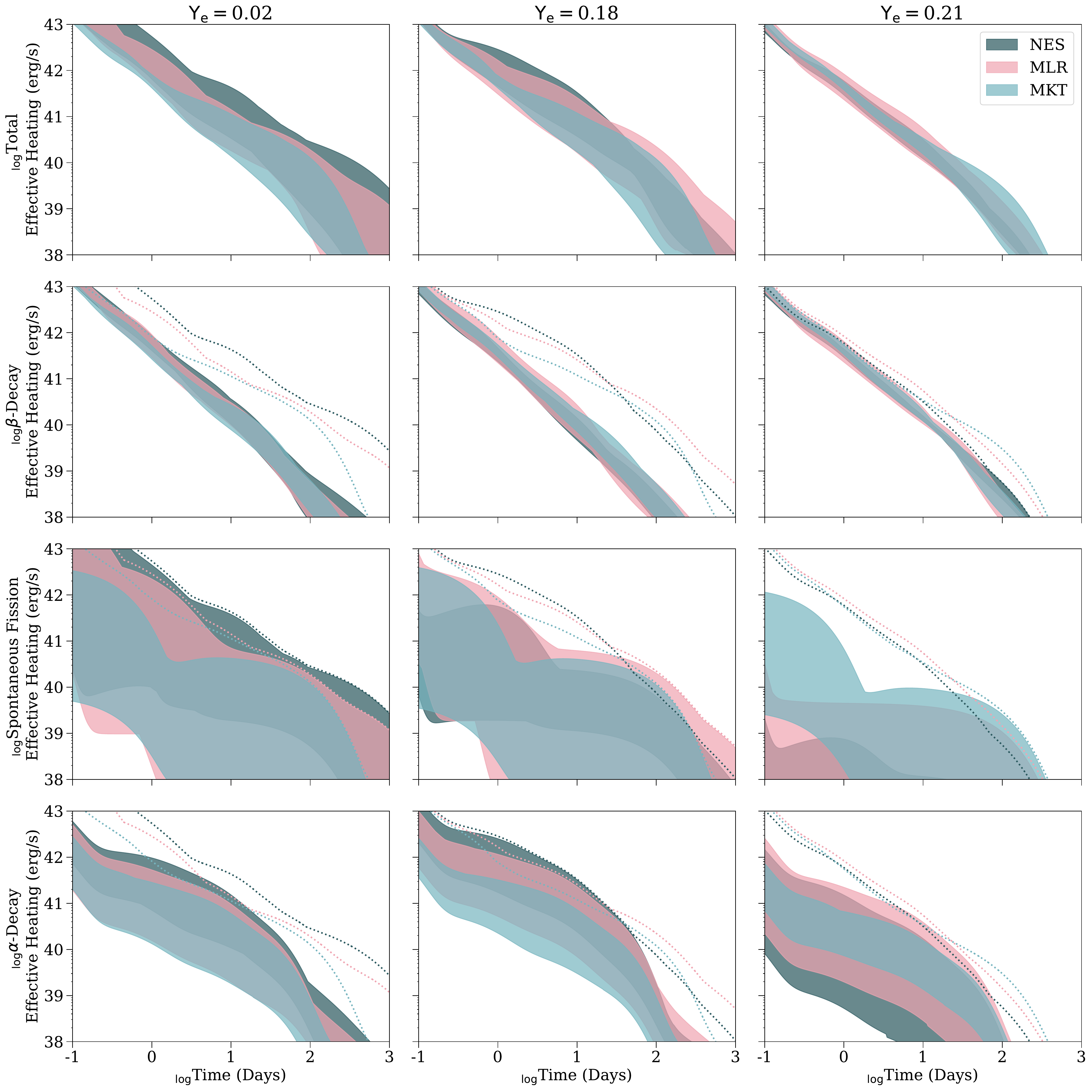}
    \caption{
    Range of effective heating rate predictions for all nuclear mass models from Table \ref{tab:masses} when a single \ye{} trajectory is considered.
    Looking at the columns from left to right, the initial electron fraction increases from 0.02 to 0.18 to 0.21. 
    The width of each band corresponds to the range of heating predictions for the NES (dark blue), MLR (pink), or MKT (light blue) simulations with the different mass models and fission yields for which we calculate heating.
    For comparison, the maximum total heating is shown for each set of simulations as a series of dotted lines.  
    }
    \label{fig:heatbands} 
\end{figure*}

The first row of Figure \ref{fig:heatbands} shows the range of total heating curves resulting from these single-\ye{} trajectories of 0.02 (left), 0.18 (center), and 0.21 (right). The width of any one shaded band comes from the use of different mass models, and corresponding fission barriers, with each band color corresponding to one set of beta decay rates from Figure \ref{fig:rateratios}. Overlap in the bands for different \ye{} cases appear as a darker region on the plot. An immediately noticeable trend is that the two lower \ye{} cases have a wider spread in the prediction of total heating than the highest \ye{} simulation does. In these lower \ye{} cases, the  NES simulations (darkest blue region) tend to show the highest \textit{total} heating rates, i.e. they provide an upper limit for the total heating. Conversely, the MKT simulations (light blue region) tend to show less total heating and therefore provide a lower limit for the same.

To explore the reason for these effects, in the next three rows of the figure, we plot the contribution to the effective heating that stems directly from beta decaying nuclei (second row), fissioning nuclei (third row) and alpha-decaying nuclei (bottom row). In all cases, the upper limit of the total heating shown in the top row is reproduced here as a faint dashed line for comparison. 
The narrow width of the beta decay heating bands, as well as their overlap, indicates that uncertainties in total effective heating, seen in the top row, cannot \textit{directly} be attributed to differences in the beta decay heating. The third and fourth rows of Figure \ref{fig:heatbands} indicate that, in fact, the largest variation in total heating instead comes from differences in the contribution of spontaneous fission and alpha decay heating. Looking at the right-most column, the \ye$=0.21$ case, we see that the total heating is dominated by beta decay, with alpha decay and spontaneous fission making up a relatively small portion of the total heating. As a consequence, the total heating for this case (top right) exhibits the least variation with different beta decay rates. At about 1 day, this variation spans only about a factor of 2. However, looking at the left-most column (\ye$=0.02$), we see that fission can substantially affect both the total heating and the uncertainty in the total heating. The choice of beta decay rates is \textit{indirectly} but strongly influencing the total heating. 

For some low \ye{} simulations, the contribution of fission is sub-dominant, but for others, fission is the majority contribution, leading to the substantial spread in the results. Alpha decay plays a similar role in the middle \ye{} cases (second column), where it largely controls the width of the band. The NES simulations contribute the upper limit in alpha decay in the middle \ye{} simulations as well as the upper limit to the heating at a day. Finally we note that the lower limits on the total heating, which are very similar between the three sets of beta decay rates, occur in simulations with minimal alpha decay and fission, and are determined primarily by the beta decay rates.

\begin{figure}[!ht]
\centering
\includegraphics[scale=0.63]{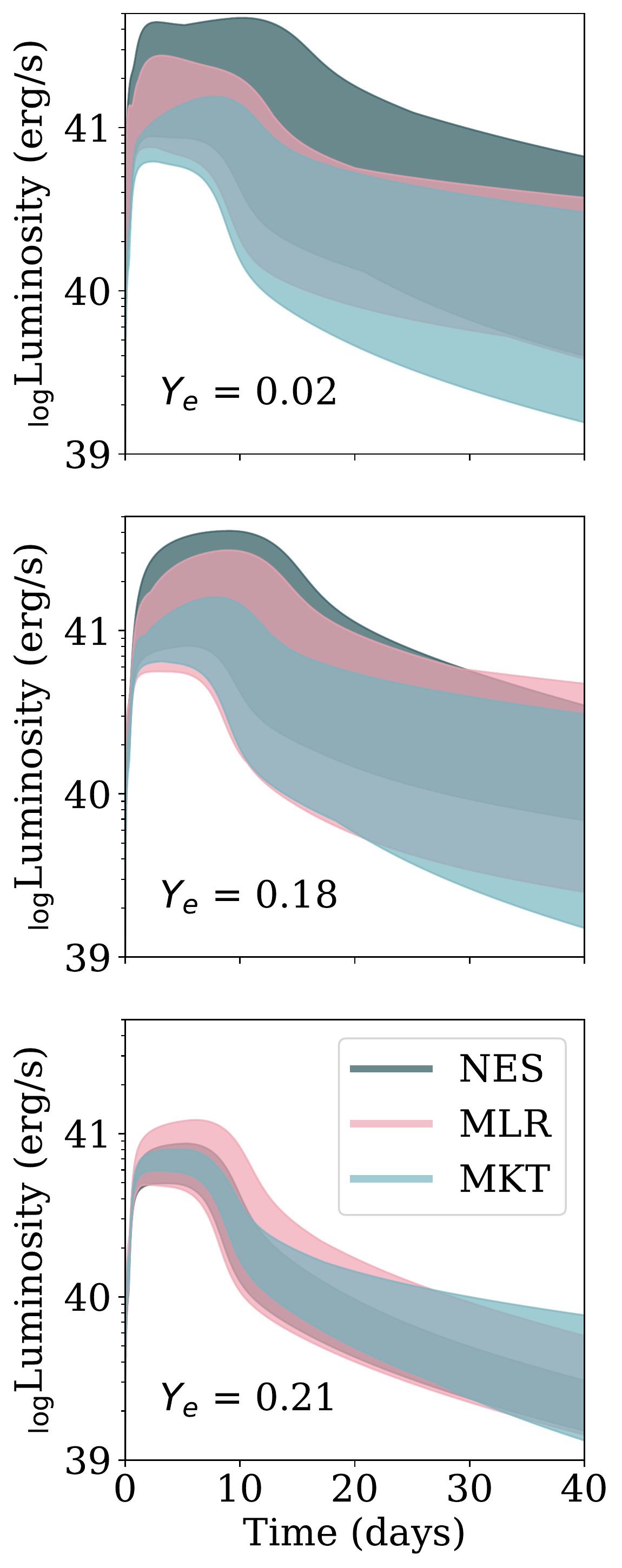}
\caption{ Uncertainty range of analytic light curve models for all theoretical nuclear models (from Table \ref{tab:masses}, when a single \ye{} trajectory is considered, with the calculation carried out to 40 days post-merger. The initial electron fraction increases from 0.02 to 0.18 to 0.21 from top to bottom. The width of each band corresponds to the range of luminosity predictions when the NES (dark blue), MLR (pink), or MKT (light blue) beta decay rates are used, and include both 50/50 and KT fission yields. 
}
\label{fig:lumband}
\end{figure}

The shape and peak luminosity of the late-time ("red") light curve are expected to be substantially influenced by the evolution of the nuclear heating (\citetalias{Zhu2021}, \citetalias{Barnes2021}). Thus, as described in Section \ref{sec:methodLC}, we compute the evolution of the light curve on a timescale of days, out to 40 days post-merger. Analogous to Figure \ref{fig:heatbands}, we show the ranges of these results in Figure \ref{fig:lumband}. We see that the beta decay rates which produce the upper limit in the overall heating produce a corresponding upper limit in the light curve. Similarly, the range of uncertainty follows the pattern of heating bands with the largest variations coming from the lowest \ye{} cases. It is interesting to note that in the \ye$= 0.21$ scenarios, the full range of variation is largely captured by the MLR rates.

While Figures \ref{fig:heatbands} and \ref{fig:lumband} show the broad uncertainty in heating and light curve evolution that can be obtained by changing the beta decay rates, they do not show the specific simulations that are sensitive to these changes. The predictions of heating, and therefore of the light curve, for some mass models are relatively insensitive to the beta decay rates. On the other hand, the predictions from other mass models show substantial sensitivity to the beta decay rates. To illustrate this point, we have separated the total heating rate results by mass model in Table \ref{tab:avgs}.

For easier comparison, we list the average ratio of total heating for NES:MLR and MKT:MLR. We also list the ratio of luminosity in parentheses, averaged over two different time periods: 1-10 days and 10-100 days. For both the average heating as well as the average luminosity, we highlight those instances where using a different set of beta decay rates results in a change of 50\% or more in bold. In the following discussion, we refer to these models with the format \mzero{}.

\begin{table*}[ht]
\centering
\begin{tabular}{|l|c|cc|cc|}
\hline
    & & \multicolumn{2}{c|}{1-10 Days} & \multicolumn{2}{c|}{10-100 Days}\\
    Nuclear Model & \ye{} & NES & MKT & NES & MKT \\
    \hline
    \multirow{3}{7em}{SLY4} & 0.02 & \textbf{1.616} (1.338) & 0.856 (0.917) & \textbf{1.583} & 0.761\\
    & 0.18 & \textbf{1.797 (1.642)} & 1.015 (1.217) & \textbf{2.662} & 0.937\\
    & 0.21 & 1.174 (1.112) & 1.469 (1.374) & 1.078 & 1.118\\
    \hline
    
    \multirow{3}{7em}{UNEDF1} & 0.02 & 1.166 (1.231) & 0.823 (0.888) & 1.182 & 1.009\\
    & 0.18 & \textbf{1.517} (1.404) & 0.557 (0.657) & 1.027 & \textbf{0.454} \\
    & 0.21 & 0.878 (0.904) & 1.167 (1.16) & 0.801 & 1.125\\
    \hline
    
    \multirow{3}{7em}{DZ33} & 0.02 & \textbf{1.674} (1.459) & 0.678 (0.82) & \textbf{2.224} & 1.089\\
    & 0.18 & \textbf{1.631} (1.416) & \textbf{0.432} (0.549) & 0.938 & 0.849\\
    & 0.21 & \textbf{0.416} (0.504) & 0.656 (0.709) & \textbf{0.384} & 1.376\\
    \hline
    
    \multirow{3}{7em}{ETFSI} & 0.02 & 1.114 (1.042) & 0.553 (0.683) & \textbf{1.779} & 1.432\\
    & 0.18 & 1.24 (1.179) & \textbf{0.333 (0.499)} & \textbf{1.799} & 0.841\\
    & 0.21 & 0.923 (0.914) & 1.241 (1.263) & 1.12 & 1.249\\
    \hline
    
    \multirow{3}{7em}{FRDM2012} & 0.02 & 1.27 (1.089) & 0.548 (0.752) & 1.438 & 0.633\\
    & 0.18 & 1.377 (1.261) & \textbf{0.453} (0.556) & 0.785 & \textbf{0.333}\\
    & 0.21 & 0.538 (0.589) & 0.642 (0.692) & \textbf{0.318} & 0.822\\
    \hline
    
    \multirow{3}{7em}{HFB22} & 0.02 & \textbf{6.591 (2.806)} & 1.215 (0.529) & \textbf{4.528} & \textbf{1.501}\\
    & 0.18 & 1.069 (0.993) & \textbf{0.412} (0.551) & \textbf{0.437} & \textbf{0.333}\\
    & 0.21 & 0.895 (0.902) & 1.177 (1.2) & 0.949 & \textbf{3.147}\\
    \hline
    
    \multirow{3}{7em}{HFB27} & 0.02 & \textbf{2.002 (1.74)} & \textbf{0.26 (0.348)} & \textbf{2.085} & \textbf{0.399}\\
    & 0.18 & 1.177 (1.083) & \textbf{0.302 (0.462)} & 0.697 & \textbf{0.299}\\
    & 0.21 & 0.962 (0.952) & 1.264 (1.27) & 1.011 & \textbf{3.158}\\
    \hline
    
    \multirow{3}{7em}{TF} & 0.02 & 1.25 (1.068)& \textbf{0.406 (0.761)} & 1.171 & \textbf{0.406}\\
    & 0.18 & 1.078 (1.11) & 0.586 (0.786) & \textbf{1.818} & \textbf{0.364}\\
    & 0.21 & 0.642 (0.719) & 0.54 (0.634) & 0.584 & \textbf{0.282}\\
    \hline

    \multirow{3}{7em}{WS3} & 0.02 & \textbf{1.549} (1.277) & 0.729 (0.856) & \textbf{1.918} & 0.976\\
    & 0.18 & 1.293 (1.253) & 0.578 (0.664) & 0.709 & 0.828\\
    & 0.21 & 0.886 (0.893) & 1.389 (1.33) & 1.131 & \textbf{4.588}\\
    \hline
\end{tabular}
\caption{\label{tab:avgs}Average of the ratios of total effective heating using NES or MKT to total effective heating using MLR beta decay rates, over different time periods. In the 1-10 days column, the average ratio of bolometric luminosity is listed in parentheses.  Bold-faced values indicate an average change of $\pm 50\%$ or more.}
\end{table*}

The heating at the later timescale of 10-100 days is in many instances dominated by the spontaneous fission of the long-lived \iso{Cf}{254}, which is a marker for actinide production. However, some models, such as the HFB models, facilitate the contribution from additional spontaneous fission heaters during this time. On the other hand, models with lower fission barrier heights tend to suppress these extra possible contributions to the fission heating, leaving only that of \iso{Cf}{254} (\citetalias{Zhu2021}, \citetalias{Barnes2021}). 

During the earlier time period of 1 - 10 days, the difference in total heating stems largely from competition between spontaneous fission, alpha decay, and beta decay heating.
This highlights the sensitivity of some mass models to both \ye{} and beta decay rates.
For example, in the case of \mone, we find that the NES simulations show approximately 60\% of the total heating coming from alpha decay by ~5 days.
Meanwhile, the corresponding MLR simulation is dominated by beta decay with a contribution from alpha decay that only rises to about 40\% by 7 days. 
Similarly, in the case of \mtwo, the MLR simulations show more than 80\% of the total heating coming from alpha decay as early as 2 days. 
When the MKT rates are used, there is still a significant contribution from alpha decay, but only up to a maximum of about 64\% around 6 days.

\begin{figure}[ht!]
\centering
\includegraphics[scale=0.26]{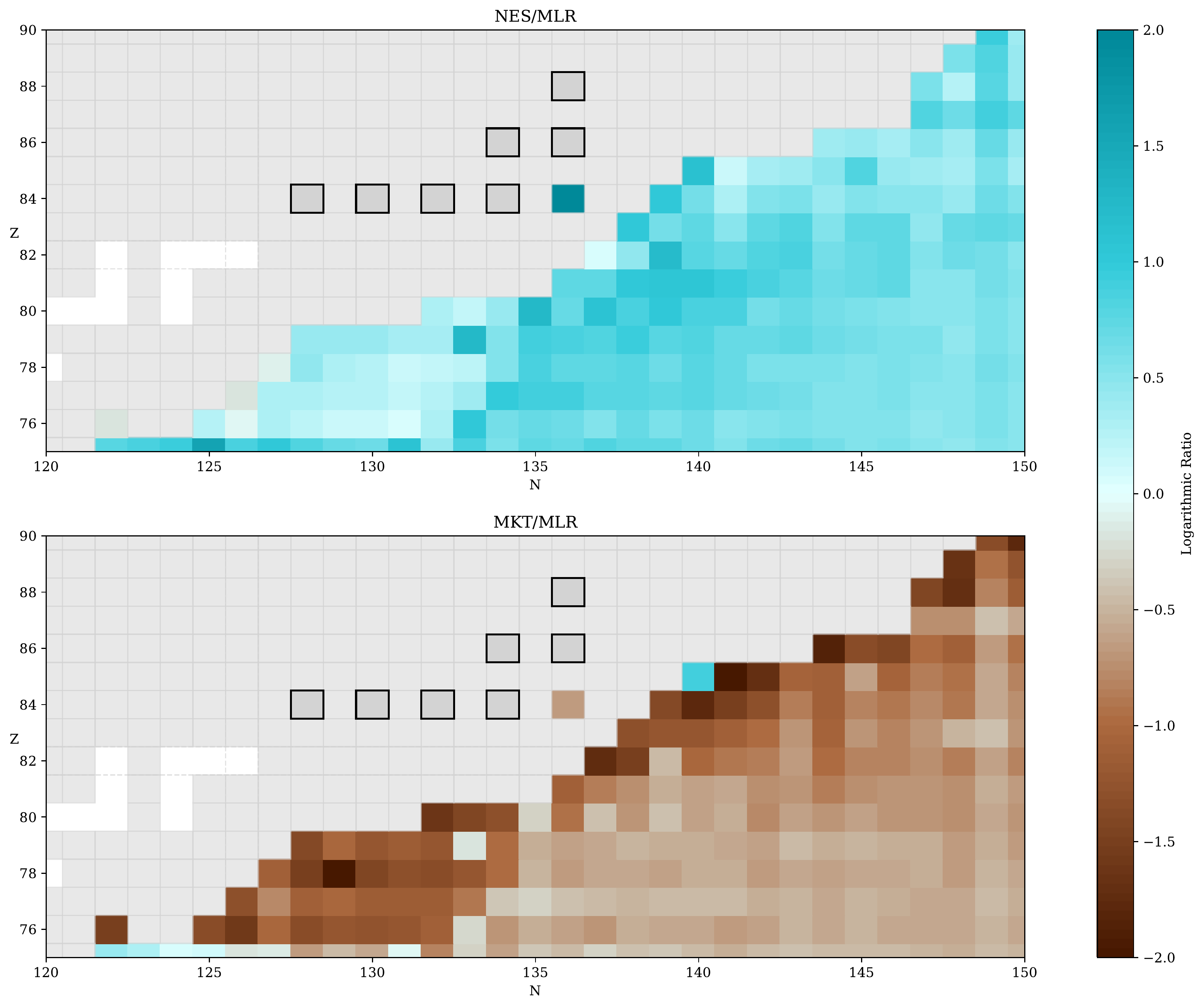}
\caption{Ratio of beta decay rates used in NES:MLR (top) or MKT:MLR (bottom) calculations in the region feeding into important alpha decay heaters (outlined in black boxes) identified in, for example, \mone. We show here only the rates subject to theoretical models; those covered by the Nubase2016 data set are shown in grey. 
}
\label{fig:alpha_zoom}
\end{figure}

In both these cases, it was the alpha decays of \iso{Po}{212}, \iso{Po}{214}, or \iso{Po}{216} that were among the top most significant contributors to the total heating. 
These lie in a region where nuclei undergo alpha decay on very short timescales and with a relatively large Q-value. 
Because of the very short timescales on which these decays occur, it is actually the populations of \iso{Ra}{224} and \iso{Rn}{222} (with half-lives of 3.6 and 3.8 days, respectively) that determine the overall contribution of the decays of their daughter polonium isotopes.
The alpha decay feeders into, for example, \iso{Ra}{224} and \iso{Rn}{222} decay on timescales that are too slow (with half-lives of 1.9 years and 1600 years,respectively) to be directly responsible for differences in heating on a timescale of a few days. 
Thus we conclude that the main source of differences lies in the unmeasured beta decay rates feeding into \iso{Ra}{224} and \iso{Rn}{222}, as well as directly into \iso{Po}{212-216} (highlighted in figure \ref{fig:alpha_zoom}); these are critical for determining the amount of material that is available for alpha decay heating, thereby determining the dominant source of total heating. 
We find that it is the cumulative effect of slight differences in the beta decay rates in the large feeder region, rather than any one specific feeder nucleus.
The use of overall slower rates (NES) feeding into this alpha decay region resulted in a large enough heating contribution from alpha decay to dominate significantly over the beta decay heating that determined the total heating (and light curve) at earlier times.

Spontaneous fission reactions occurring on time scales of days have the largest potential to make a significant difference in the overall heating, as well as the light curve, due to the large Q-values involved as well as the high thermalization efficiency of the reaction products.
We find that, especially in the cases using the HFB theoretical nuclear models, spontaneous fission heating has the potential to dominate the total heating as early as 1 day post-merger.
The enhanced heating seen in \mthree, \mfour, and \mfive, for example can be attributed largely to differences in the predicted spontaneous fission heating rates. 
However, there is still a great deal of variety.

\begin{figure*}[ht!]
    \includegraphics[width=\textwidth]{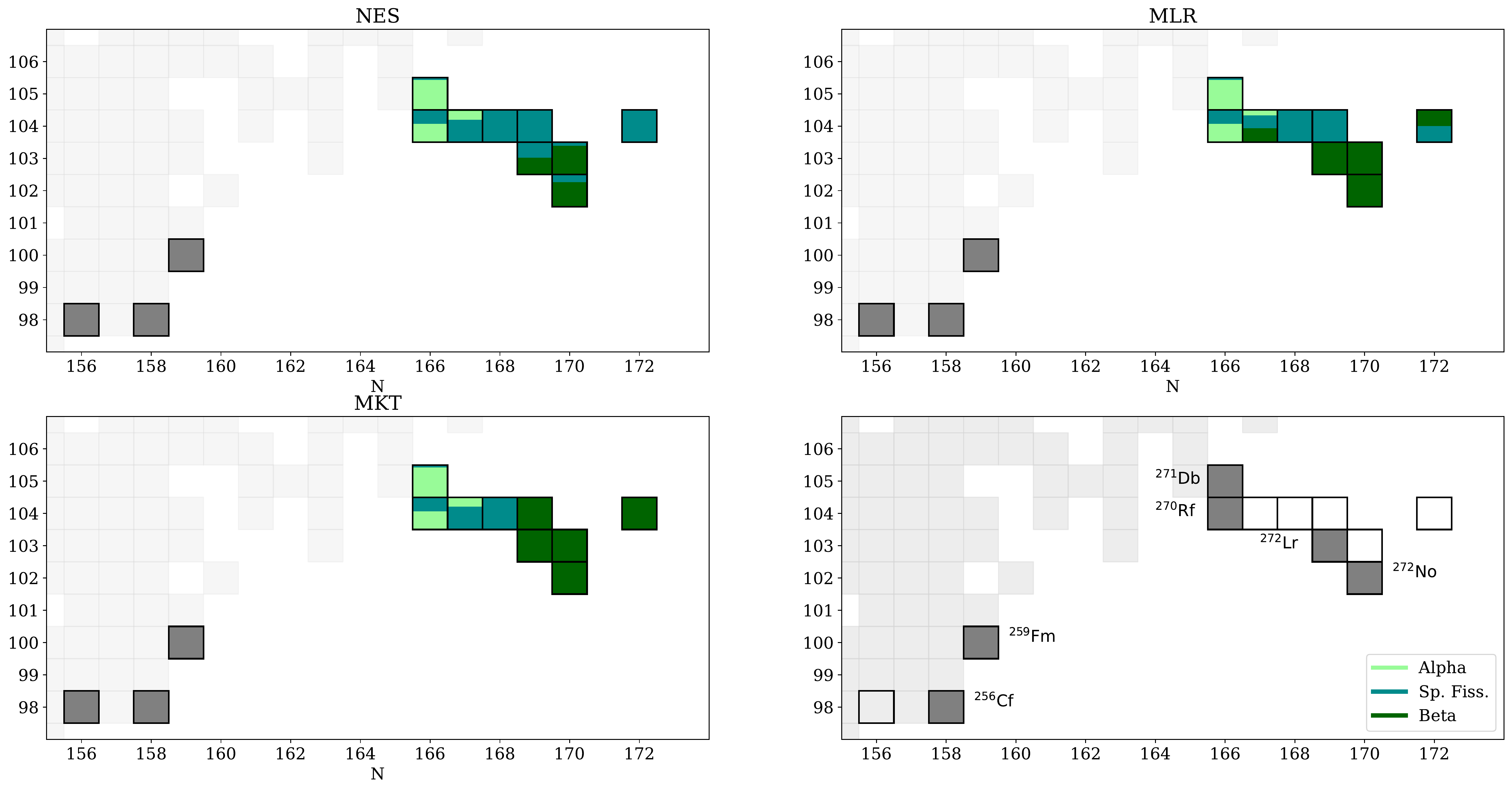}
    \caption{Theoretical branching ratios for a selection of key nuclei that have the potential to be important spontaneous fission heaters, as identified in, for example, \mthree. 
    The grey-filled isotopes have experimentally determined values; the colored isotopes remain unmeasured. 
    Light green, teal, and dark green represent the branchings expected for alpha decay, spontaneous fission, and beta decay, respectively.
    In the bottom right panel, we show the chemical names for highlighted isotopes for ease of identification.}
    \label{fig:br_fission} 
\end{figure*}

By one day post-merger, \mthree{} shows approximately $77\%$ of the total heating as coming from spontaneous fission heating, compared to only $33\%$ at the same time in the corresponding MLR simulation. 
By three days, the spontaneous fission heating in the MLR simulation loses out to beta decay heating, while the NEs simulation shows it continuing to dominate the total heating out past ten days.
There are two mechanisms largely responsible for this behavior.
One involves directly competing theoretical branching ratios for potential fission heaters. 
The colored regions of figure \ref{fig:br_fission} indicate where this occurs, and show the theoretical branching ratios for alpha decay, spontaneous fission, and beta decay.
We found that the isotopes \iso{No}{272} and \iso{Lr}{271} appeared to consistently be responsible for a large part of the total heating in NES simulations.
The significantly slower beta decay rates predicted in these cases allowed for the spontaneous fission mechanism to compete with beta decay. 
Contrarily, the beta decay rates predicted in MLR and MKT are fast enough to yield a beta decay branching ratio of almost 100\%.

The isotopes of Rutherfordium (Z=104) appeared to contribute to different degrees in the NES and MLR calculations, yet were not among the top heaters in the MKT calculation. 
In NES calculations, the heavier isotopes (N=168,169,172) appeared to contribute the most to the total heating at early times. 
For example, the spontaneous fission of \iso{Rf}{273} alone was responsible for approximately 28\% of the total heating at on day post-merger.
On the other hand, it was the "lighter" isotopes (N=166,167) that contributed the most to the total heating in MLR calculations, and did so for a more extended period of time in both MLR and NES calculations.

We attribute this to the second mechanism responsible for differences in the role of fission in our calculations: differences in feeder decay chains that build up different abundances available for decay. 
One very obvious example of this is that of \iso{Db}{271}. 
This isotope only appears in the MLR calculation, as this is the only one in which decay into it is allowed. 
Furthermore, in the MLR calculation in which it appears, the heating from its decay via spontaneous fission competes with or even exceeds that of alpha decay, despite a very small fission branching.

This difference in populations also affects the extent to which fission heaters with measured rates are able to contribute to the total heating. 
These are also highlighted in figure \ref{fig:br_fission}, and include \iso{Cf}{254}, \iso{Cf}{256}, and \iso{Fm}{259}. 
In the MKT calculation, the population of \iso{Fm}{259} is blocked via beta decay. 
The population of its alpha decay feeder, \iso{No}{263}, is also blocked via beta decay, resulting in \iso{Fm}{259} not being able to contribute significantly to the heating in MKT calculations.
However, in NES calculations, enough material is able to decay into \iso{Fm}{259} such that its contribution to the spontaneous fission heating is significant; alone it is responsible for roughly $38\%$ of the total heating at 2 days post-merger. 

While these calculations represent the results obtained by using a single-\ye{} trajectory, they highlight the influence that the choice of mass model can have when combined with \ye{}. 
The use of certain mass models with a given \ye{} unlocked a wide variety of potential heaters that impacted the evolution of the light curve.
We emphasize that experimental data for several of these unmeasured isotopes would prove highly valuable in constraining this uncertainty. 

\subsection{Combined Trajectories}

\begin{figure*}[ht!]
    \includegraphics[width=\textwidth]{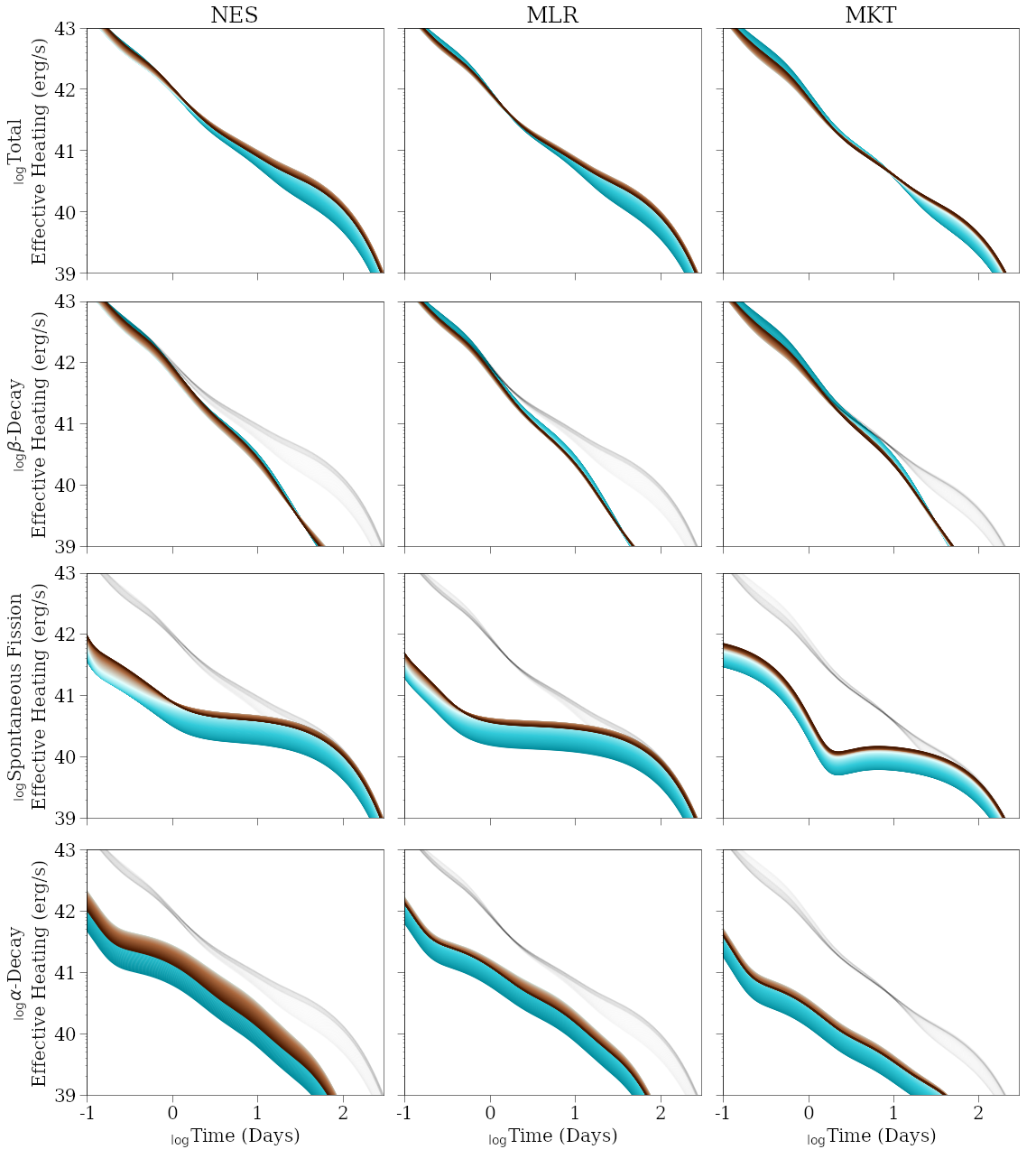}
    \caption{Range of effective heating rate predictions of FRDM2012 mass model when a composite-\ye{} trajectory is considered.
    From left to right, the beta decay description used is NES, MLR, MKT. 
    The color scheme corresponds to that shown in Figure \ref{fig:ye_dist}.
    As in Figure \ref{fig:heatbands}, we replicate the total heating curves shown in the top row in the bottom three rows as light grey lines for easier comparison.}
    \label{fig:heatyedist} 
\end{figure*}

\begin{figure*}
\includegraphics[width=\textwidth]{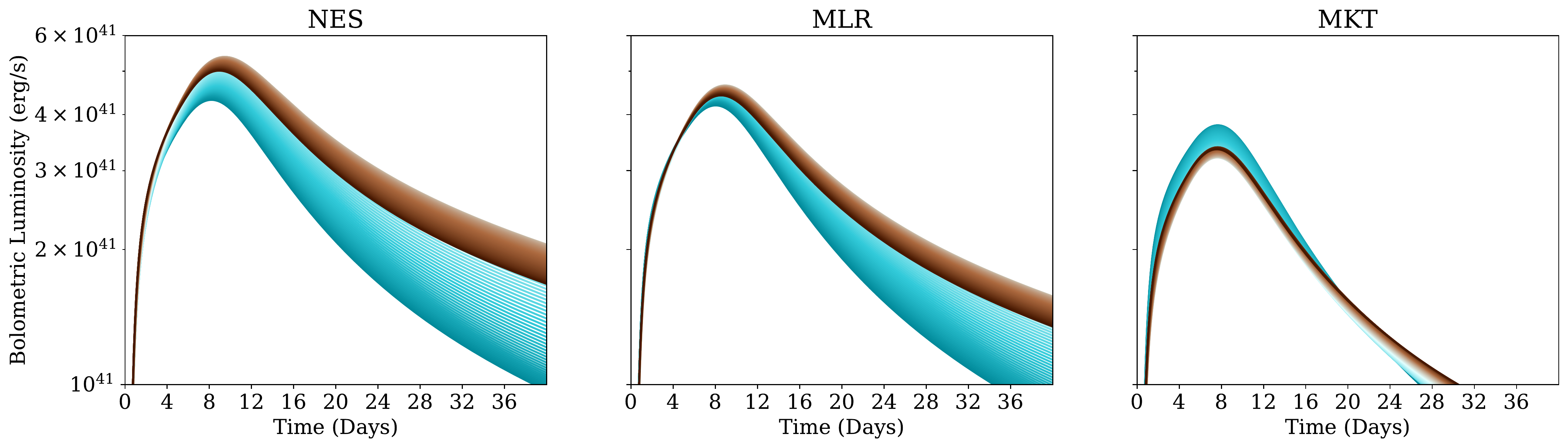}
\caption{Uncertainty range of analytic light curve models for FRDM2012 nuclear model with multiple-\ye{} composition. From left to right, the beta decay description used is NES, MLR, MKT. 
The color scheme corresponds to that shown in Figure \ref{fig:lumyedist}, with the most striking blue representing the behavior of the most neutron-poor material and the darkest brown that of the most neutron-rich material.}
\label{fig:lumyedist}
\end{figure*}

Since it is expected that element synthesis will occur in outflows with a range of electron fractions, we now turn to the variation in outcomes that is produced by employing different sets of beta decay rates in scenarios with multiple trajectories but using a single mass model.  Accordingly, Figure \ref{fig:heatyedist} shows the nuclear heating obtained from the linear combinations of individual \ye{} trajectories (combined trajectories) using \textit{only} the FRDM2012 mass model, as described in section \ref{sec:models}.
The coloring of the individual lines corresponds to those in Figure \ref{fig:ye_dist}, i.e. more brown representing a combination weighted toward low-\ye{} and bluer representing one weighted toward high \ye{}.
The left, center, and right columns represent NES, MLR, and MKT simulations, respectively.  Note that in all cases we have a substantial fraction of high \ye{} (\ye$ > 0.2)$ material that has little to no fission.

As in Figure \ref{fig:heatbands}, the top row of Figure \ref{fig:heatyedist} shows the total heating for each of the combined trajectories.
Similarly, the second, third, and fourth rows show the individual contributions from beta decay, spontaneous fission, and alpha decay heating, respectively and the total heating is shown as faint grey lines for comparison.
We find the heating out to at least one day is dominated by beta decay for all three beta sets of beta decay rates. 
Furthermore, the beta decay heating is roughly independent of the combined trajectory, as can be see from a comparison of panels in the second row.

As mentioned, each of the composite trajectories has a significant amount of high-\ye{} material and this  can "dilute" the heating contributions from spontaneous fission and alpha decay.
Indeed, the spontaneous fission heating does not appear to significantly dominate the shape of the total heating until tens of days, for any of the combined trajectories. 
We find the most potential for early-time (order days) contribution to the total heating from spontaneous fission in the low-\ye{} weighted NES simulations.
Similarly, the effect of the alpha decay heating is diluted enough that for no combined trajectory does it ever dominate the total heating.
However it is apparent from Figure \ref{fig:heatyedist} that the description of the total heating is not complete without accounting for both the spontaneous fission as well as the alpha decay heating.

In Section \ref{sec:single}, we saw the largest contribution from spontaneous fission in the lowest \ye{} case. 
We see this same behavior in the third row of Figure \ref{fig:heatyedist}.
We see the combined trajectories with the largest proportions of the lowest-\ye{} material showing the most spontaneous fission heating.
As a larger proportion of high-\ye{} material is included, we begin to see this dilution effect, and the amount of spontaneous fission heating decreases accordingly. 
Section \ref{sec:single} also showed the largest amount of alpha decay heating in the "semi-neutron rich" (\ye{} of 0.18) case.
This behavior is reflected in the fourth row of Figure \ref{fig:heatyedist}, and is most apparent for the NES simulations. 
There is an increase in the contribution of alpha decay heating as more neutron-rich material is included until a point where the material becomes too neutron-rich and material is more efficiently deposited into the higher-Z fissioning region, thus contributing less to the alpha decay heating. 

In the case of both spontaneous fission and alpha decay, the point in time at which all these effects occur depends on the beta decay rates.
For example, the number of days after the merger at which beta decay no longer closely approximates the total heating occurs sooner in NES simulations than MLR, which in turn occurs sooner than in MKT simulations.
We find that the contribution of alpha decay heating has the most potential to be significant in NES simulations that are more heavily weighted toward the middle of our \ye{} range.
There is also a more significant difference between the potential for significant alpha decay heating for mid-\ye{} combined trajectories compared to those weighted toward low \ye{} for NES simulations than both MLR and MKT simulations. 
In these latter cases, there is more similarity between low- and mid-\ye{} weighted combined trajectories. 
This is consistent with the result obtained in Section \ref{sec:single}, where we found the largest contribution to alpha decay heating to be in the \ye$=0.18$ case in NES simulations. 

We find that the use of a different set of beta decay rates changes the time scales on which these differences appear, and the extent to which they affect the total heating.
However, when these differences are propagated through the light curve calculations, we find that the differences are more subtle.
We show the light curves resulting from the combined trajectory effective heating results in Figure \ref{fig:lumyedist}, following the same coloring convention as the heating.
The left, middle and right panel show the light curves for NES, MLR and MKT simulations, respectively. 
In all three cases, we find that the overall shape of the light curves are consistent for all three sets of calculations, e.g. there are no plateaus or bumps present in some but not others.

The most apparent difference in the light curves lies in the behavior after approximately 4 days.
The NES and MLR simulations show similar behavior with the NES simulations yielding a higher peak luminosity for mid to high-\ye{} weighted combined trajectories.
In addition to a smaller peak magnitude, the low-\ye{} weighted combinations show an earlier peak than the more neutron rich trajectories. 
Furthermore, these tend to decay more quickly.
The MKT calculations yield slightly different results, with the lowest-\ye{} material yielding the largest peak magnitude.
This is consistent with the total heating behavior observed in the MKT simulations, shown in the top row of Figure \ref{fig:heatyedist}.
We attribute this to the dominant heating mechanisms in MKT calculations being beta decay for a longer period of time, which is largest in the higher-\ye{} weighted trajectories. 
Thus the alpha decay heating contribution, which is largest for low-\ye{} dominated trajectories, is unable to compete until later in time. 
This point is reflected in the MKT light curves, in which there is a flip, and the low-\ye{} trajectories have the highest luminosity. 

\section{Applications for Nuclear Cosmochronometry}\label{sec:cosmoresult}

\begin{figure}
    \centering
    \includegraphics[width=0.4\textwidth]{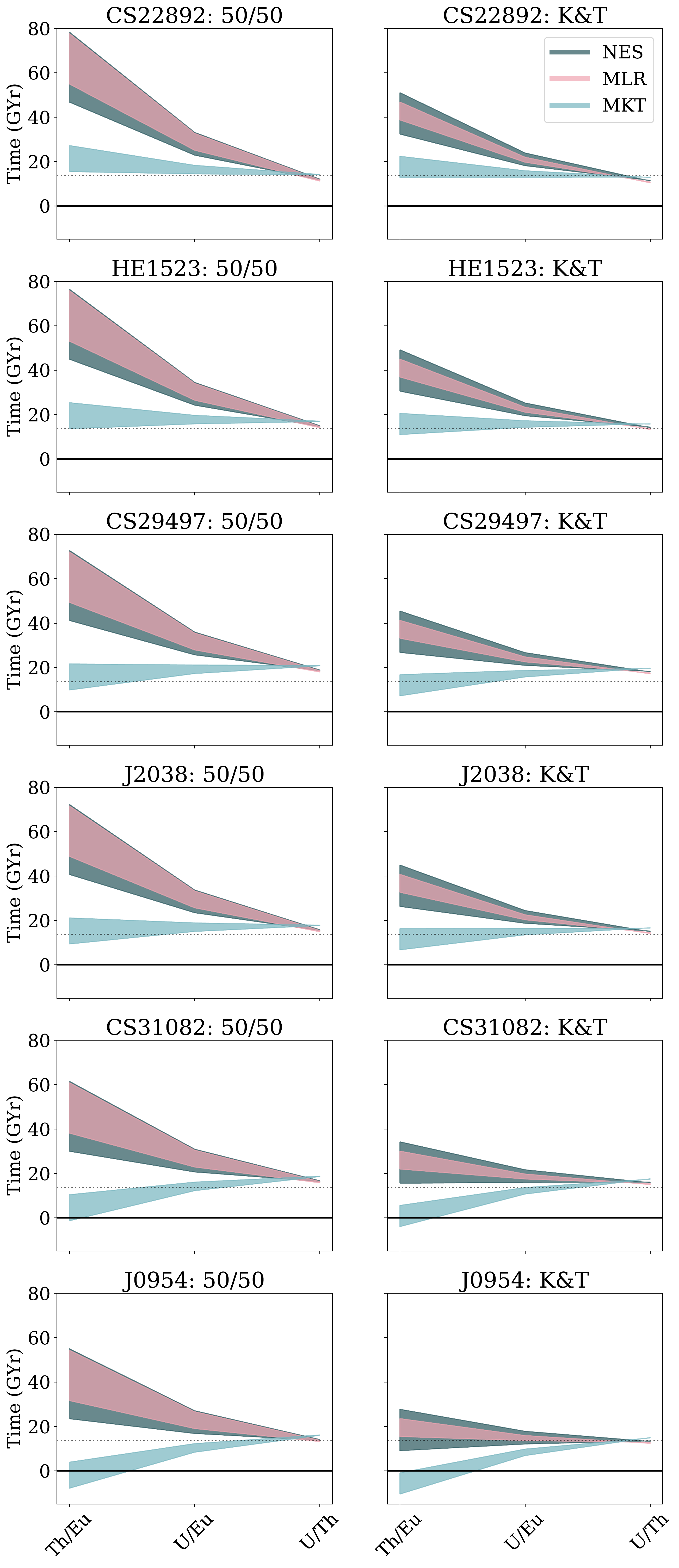}
    \caption{Maximum uncertainty from the use of different beta decay rates for the stars listed in Table \ref{tab:stars}, with each row corresponding to a different star. 
    The left column shows ages computed using the 50/50 fission yields, while the right column contains those obtained using the K\&T fission yields. 
    As throughout this work, NES calculations are shown in dark blue, while MLR and MKT calculations are shown in pink and light blue, respectively. 
    The regions are filled in to show the range of values obtained using the three different chronometers, as described in Equations \ref{eq:ThEu}-\ref{eq:UTh}. 
    Dashed horizontal lines indicate a value of 13.7 GYr; solid horizontal lines indicate a value of 0 GYr.} 
    \label{fig:ages_maxmin}
\end{figure}

Using the same combined trajectories described in Section \ref{sec:models} (which all use the FRDM2012 mass model), we calculate the ages \footnote{We use "age" as a concise way to refer to the time since the enrichment event- not \textit{necessarily} the time since the formation of the star itself.} of several stars, as described in Section \ref{sec:cosmo}. 
We show the range of results of these calculations in Figure \ref{fig:ages_maxmin}.
We emphasize that the choice of mass model will quantitatively influence the results shown in Figure \ref{fig:ages_maxmin} and the analysis here is presented only for FRDM masses. 
The left and right columns of Figure \ref{fig:ages_maxmin} use the 50/50 and K\&T fission yields, respectively.
To obtain these ranges, we use the quoted observational values in Table \ref{tab:stars} for $\rm log_\epsilon(X/Y)_{obs}$ and the theoretical abundances from each combined trajectory for $\rm log_\epsilon(X/Y)_0$.  
For the purposes of Fig. \ref{fig:ages_maxmin}, we have not included observational uncertainties. 
We will do so later in this section.

Global beta decay rate predictions are a critical component of cosmochronomotery studies.  
For example, the use of theoretically calculated lanthanide \textit{and} actinide abundances can yield age estimates that show a substantial spread depending on the choice of beta decay rates \citep{Holmbeck2018}. 
We show this effect using our calculations in Fig. \ref{fig:ages_maxmin}.
We show the maximum uncertainty in age for the six stars in table \ref{tab:stars} from NES (dark blue), MLR (pink), and MKT (light blue) calculations. 

\begin{figure}
    \centering
    \includegraphics[width=0.45\textwidth]{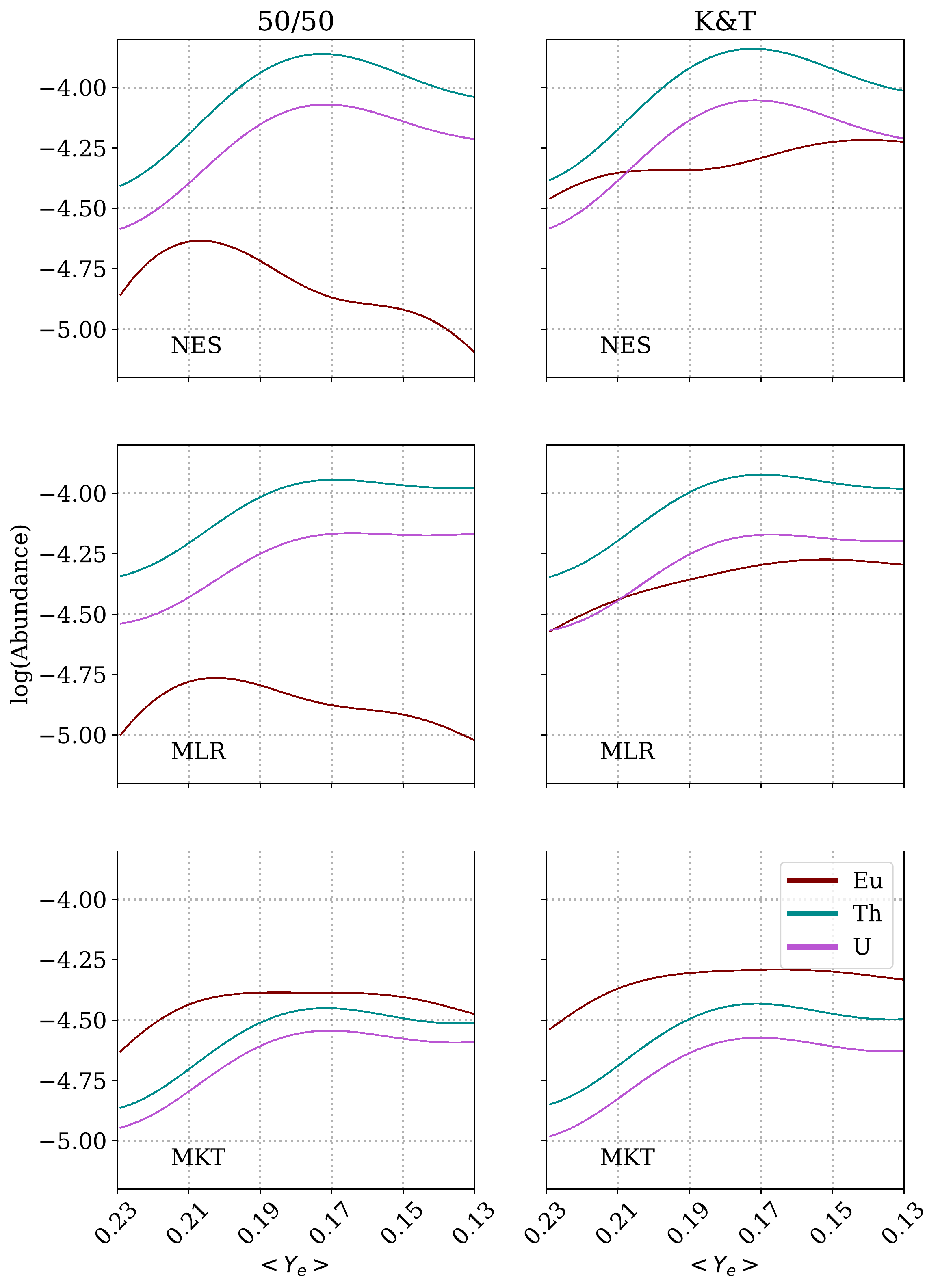}
    \caption{Composite values of initial production abundances of europium (red), thorium (teal), and uranium (pink), separated out by beta decay rates used. NES, MLR, and MKT simulations are shown in the top, middle, and bottom rows, respectively. The left column shows results obtained using the 50/50 fission yields, and the right column shows those obtained using the K\&T fission yields. The horizontal axis shows average \ye{} (with increasing neutron richness) of the combined trajectories.}
    \label{fig:diagnostic}
\end{figure}

Use of the rare earth - actinide chronometer pairs with  NES or MLR simulations tends to produce ages estimates that are high relative to the actinide-only chronometer.
This effect is most pronounced in actinide deficient stars, as can be seen from top row of Figure \ref{fig:ages_maxmin}.  
We attribute this to an overproduction of actinides relative to rare earths, as shown in Fig. \ref{fig:diagnostic}, where we have plotted the composite, i.e. for the combined trajectories, abundances of europium (red), thorium (teal), and uranium (pink).  
The overproduction is largest for the 50/50 fission yields.  
This effect is ameliorated with the use of the K\&T fission yields since this model spreads out the fission daughter nuclei over a larger range of mass number.
In all cases, for a more actinide deficient star, the amount of time that is necessary for the overproduced actinide content to decay to match observed abundances is larger.
We see this reflected in Figure \ref{fig:ages_maxmin}, where the stars are sorted in order of ratio of increasing actinide to rare earth abundance.

In contrast, the use of MKT beta decay rates in some astrophysical conditions yields theoretical initial production (Th/Eu) values that are lower than those observed in some stars.
This results in lower ages when Equations \ref{eq:ThEu} and/or \ref{eq:UEu} are applied, as compared with the actinide-only chronometer.  
As can be seen in Figure \ref{fig:diagnostic}, MKT tends to produce simultaneously less actinides and more rare earths than do other theoretical formulations of beta decay rates.  
This effect is most pronounced with the actinide boost star, J0594, as well as stars at the high end of actinide normal, e.g CS31802.  
Again, the effect is mitigated slightly when using the more diverse fission daughter product distribution (K\&T), which increases somewhat the predicted europium yield, as well as the yields between $57<Z<63$.
Indeed, europium production in K\&T simulations shows less sensitivity to beta decay rates than simulations with the 50/50 fission daughter product distribution. 
The lower actinide population in MKT calculations is consistent with previous calculations (\citetalias{Zhu2021},\citetalias{Barnes2021}), where the relatively fast MKT beta decay rates above the $\rm{N=126}$ shell closure (which can be seen in figure \ref{fig:rateratios}) were found to inhibit the buildup of a significant actinide population.
 
\subsection{Actinide Constraint}
\begin{figure}
    \centering
    \includegraphics[scale=0.36]{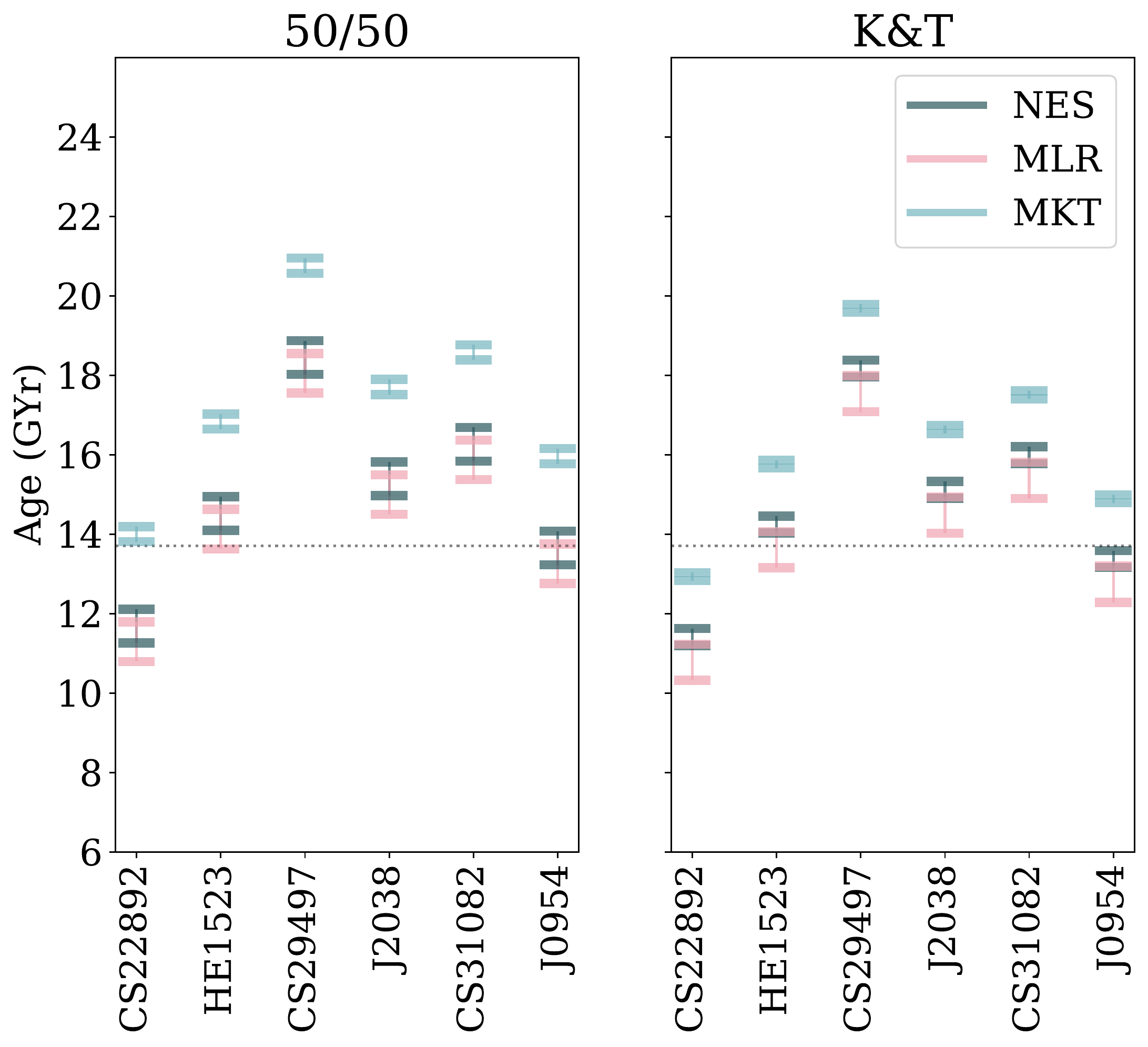}
    \caption{Uncertainty in age estimates due to uncertain nuclear physics and \ye{} for various stellar observations. The ages in this figure are calculated solely using the U:Th abundance.
    A horizontal dotted line indicates the value of 13.7 GYr.}
    \label{fig:ages_uncertainty}
\end{figure}

While the actinide:lanthanide ratios yield a large spread of results, thorium and uranium are generally produced concomitantly, resulting in smaller uncertainties.
In Figure \ref{fig:ages_uncertainty} we focus on this uncertainty, obtained solely from using variation in the theoretical values of $\rm{log_\epsilon (U/Th)}$ in our models.
Each colored bar corresponds to the range of results we obtain using NES (dark blue), MLR (pink), and MKT (light blue) simulations, along with the 50/50 (plotted on the left) and K\&T (plotted on the right) fission yields.
We find from this figure that there is more star-to-star variation, than variation from the use of different beta decay rates, fission descriptions, or combined trajectory sets.

We also find here that similar actinide abundances are produced in NES and MLR simulations, while MKT shows consistently different behavior. 
The NES and MLR simulations produced comparable amounts of thorium as well as uranium, with NES simulations never exceeding a factor 0.8-1.3 times the corresponding MLR simulation abundance.
The result of this is largely overlapping age estimates stemming from the actinide chronometers, as seen in Figure \ref{fig:ages_uncertainty}.

In comparison, the MKT simulations yielded roughly only one third the actinide abundances compared to MLR. 
However, the difference between thorium and uranium production within MKT simulations was small, as can be seen in the bottom panel of Figure \ref{fig:diagnostic}. The overall effect of this translates into larger age estimates for MKT simulations, as shown by the light blue error bars in Figure \ref{fig:ages_uncertainty} being consistently centered at larger values than the pink or dark blue.  We point out, though, that the ratio is not necessarily smaller because either uranium or thorium specifically is less effectively produced.
Rather both are inefficiently produced yielding overall smaller abundances.

\subsection{Chronometric Agreement}
\begin{figure}
    \centering
    \includegraphics[scale=0.36]{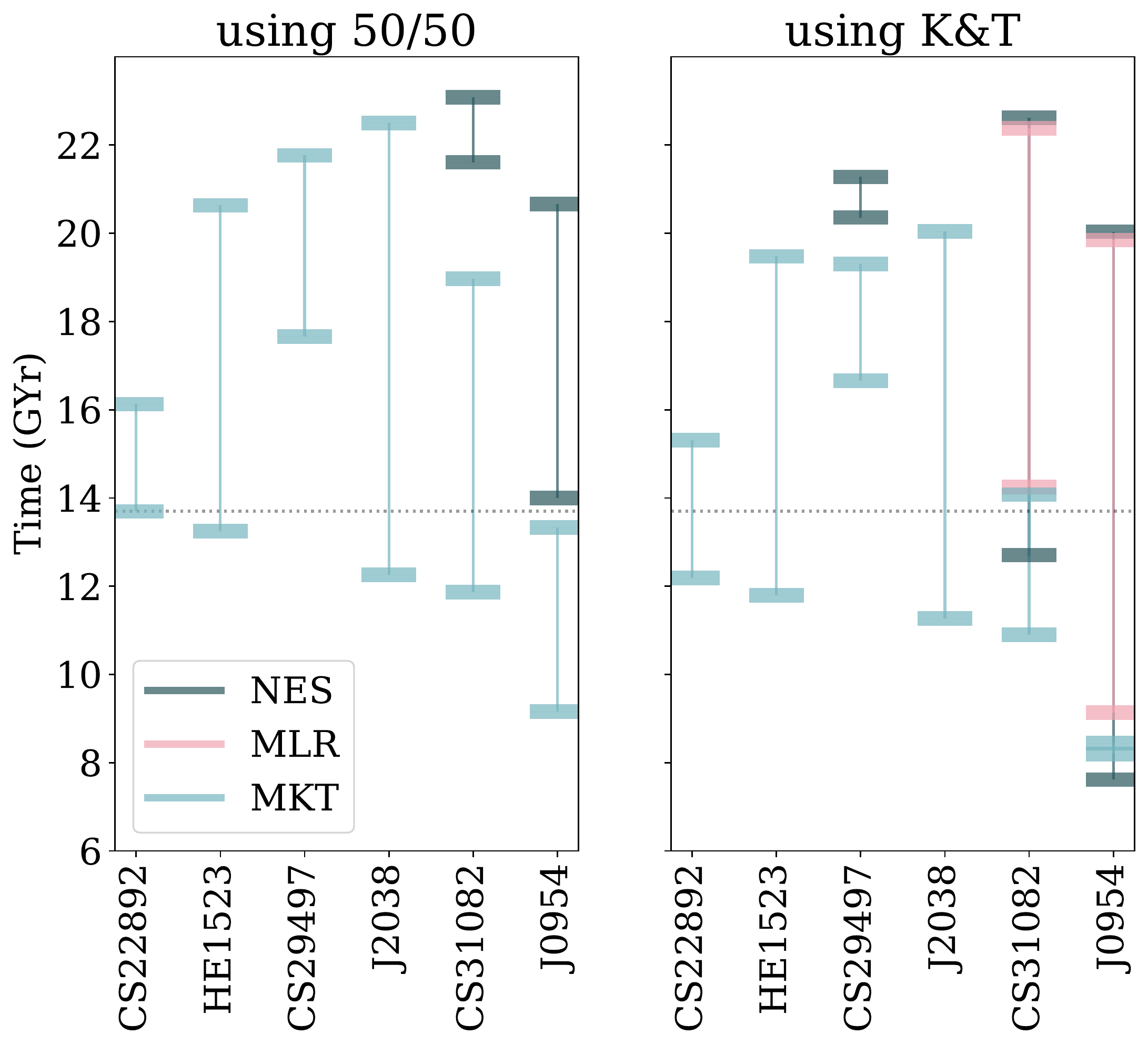}
    \caption{Full range of predicted age values when the full range of observational error bars plus the full uncertainty from the composite trajectories are used. For the values plotted in this figure, the age predictions made by all three chronometer ratios agree to within 1\% of their average.}
    \label{fig:bigloop_errorbars}
\end{figure}

An important assumption in Equations \ref{eq:ThEu}-\ref{eq:UTh} is that of a single enrichment event, i.e. that the lanthanide and actinides observed in the r-process enhanced star stem from the same event which occurred at time $\rm{t_0}$.
Beyond this, the chronometry equations make no other assumptions and are derived from the nuclear decay equation. 
Hence if the r-process elements in a given star come from a single event and if the abundances from this event have been correctly predicted, then the three chronometers should provide the same age estimates within observational uncertainty.

With this in mind, we begin with our model set which includes all the combinations of trajectories shown in Figure \ref{fig:ye_dist}, each computed for all three different beta decay rates.  
We then select only those models for which each of the chronometers yield the same age within the quoted observational error bars, terming this "chronometric agreement" (or simply "agreement").  
We show the results of this procedure in Figure \ref{fig:bigloop_errorbars}.  
We see that for many stars no chronometric agreement exists for our selection of combined trajectories with the MLR and NES rates, consistent with the results of Fig. \ref{fig:ages_maxmin}. 
However we caution that all our simulations for this analysis were performed with the FRDM mass model, and this conclusion may change when a wider variety of theoretical predictions for off-stability masses are considered. Furthermore, we have used only the combined trajectories from Fig \ref{fig:ye_dist}. Other combinations, particularly those weighted towards even higher values of \ye{} (or simply more heavily weighted toward the higher-\ye{} range of our selection) could produce agreement for these stars.

\begin{figure}
    \centering
    \includegraphics[scale=0.3]{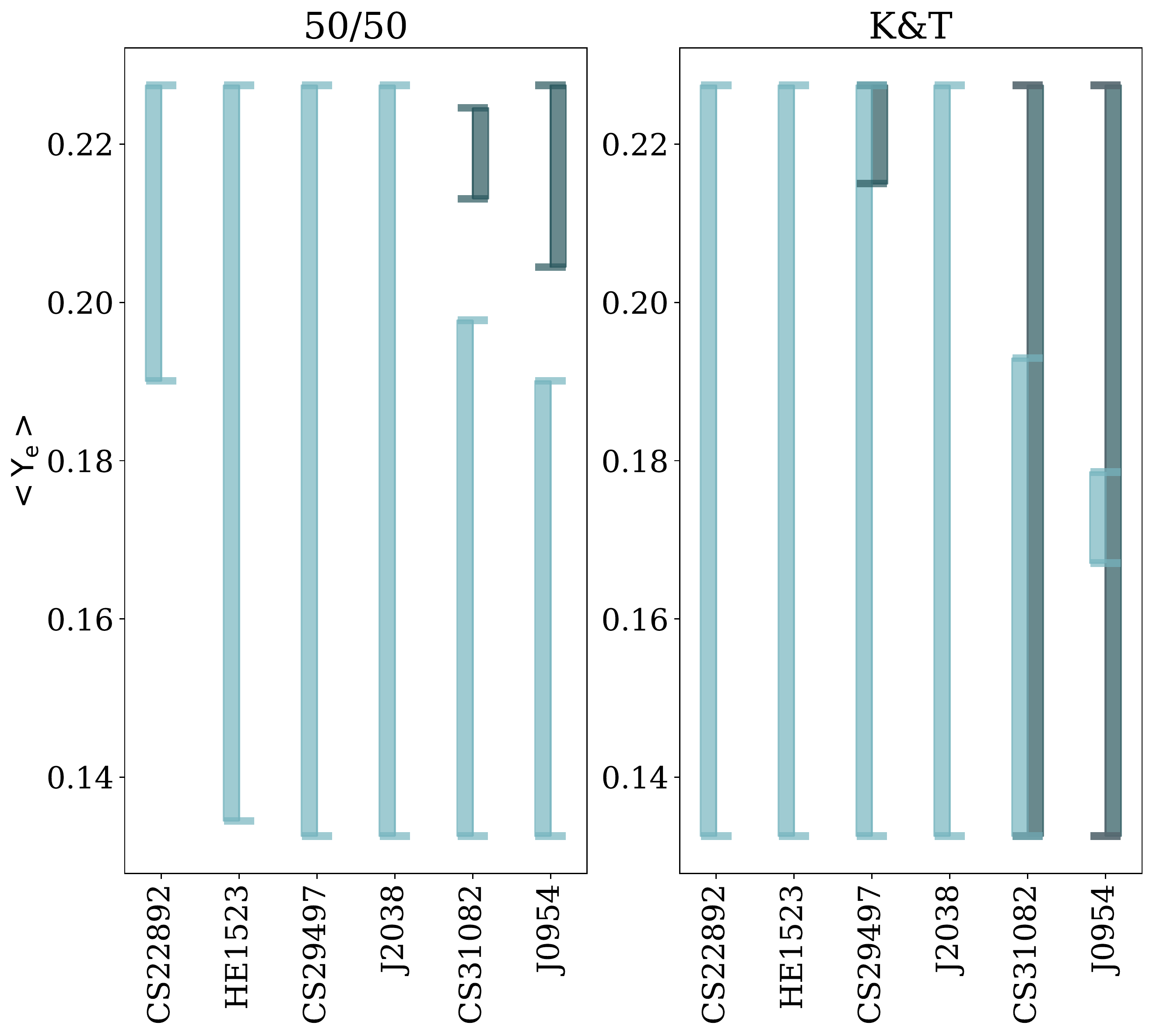} 
    \caption{Combined trajectories (as described in Section \ref{sec:method}, represented on the vertical axis by $<\rm{Y_e}$) which yield chronometric agreement for each of the stars in our sample. In this figure we show only those values obtained with NES (dark blue) and MKT (light blue) calculations, but note that there was overlap in the case of NES and MLR for CS31082 and J0954 when the K\&T fission yields were used.}
    \label{fig:which_comb}
\end{figure}

On the other hand, for all stars there are some simulations with the MKT rates that produce chronometric agreement, again due to their more limited production of actinides as compared to the rare earths.  
For each star the range of ages that are in agreement is larger in Fig. \ref{fig:bigloop_errorbars} than in Fig. \ref{fig:ages_uncertainty}.  
This is because we have taken into account observational uncertainty in the former.

It is interesting to see what sort of distribution of electron fraction is needed to produce the chronometric agreement. 
We show instances in which this occurs in figure \ref{fig:which_comb} for NES (dark blue) and MKT (light blue) calculations. 
As is consistent with Figure \ref{fig:bigloop_errorbars}, we find that some or all of the MKT calculations yield chronometric agreement when observational error bars are taken into account. 
In the case of the most actinide poor star, CS22892, all MKT calculations agree when the K\&T fission yields are used.
This changes when the 50/50 yields are used, and only those combined trajectories weighted towards high \ye{} yield agreeing results. 
Contrarily, for the stars with the largest actinide enhancement (CS31082 and J0954), MKT calculations show agreement with those trajectories weighted towards mid and/or low \ye{}.

For CS31082 and J0954, there is overlap in the NES and MLR calculations that agree when the K\&T fission yields are used.  
This is consistent with previous work that showed agreement with K\&T fission yields for the J0954 using the MLR and FRDM combination as long as the average $Y_e$ was sufficiently high, and the actinides where "diluted" \citep{Holmbeck2018}.
However, if the fission daughter product distribution is taken to be 50/50, none of the  MLR trajectories we have considered yield agreement, as do only a selection of NES calculations with trajectories weighted towards high \ye{}. This is in stark contrast to the aforementioned behavior of the MKT calculations, which favor low \ye{} weighted trajectories. This contrast is consistent with the abundance patterns resulting from these calculations, as seen in Figure \ref{fig:diagnostic}.
NES calculations more effectively reproduce a large actinide population, to the point of overproducing actinides.
Thus only a small amount of low \ye{} material is sufficient to produce a large actinide abundance.

\section{Conclusion}\label{sec:conclusion}
We performed a targeted study to investigate the impact of global decay rates on key aspects of \rp{} nucleosynthesis and kilonova modeling. 
We combined three sets of beta decay rates with nine different mass models and two fission daughter product distributions for our nucleosynthesis calculations.
Furthermore, we considered three single-\ye{} trajectories for the full suite of nuclear inputs in order to probe the role of fission heating in our calculations.
We also considered several ensembles of trajectories for a subset of nuclear inputs.
We compared the abundances obtained 1 Gyr post-merger for these ensembles to astronomical observations for six \rp-enhanced metal-poor stars.

For the single trajectory cases, we found a substantial difference in the predicted total heating from different mass models. The magnitude of this difference was sensitive to the value of $\rm{Y_{e}}$, and almost entirely due to differences in the predicted alpha decay and spontaneous fission heating.
This was especially the case in trajectories with initial fraction at or below 0.18, as this was where a significant amount of fission or alpha decay could occur.
We provided a closer investigation of some instances where the change in beta decay rate translated to an increase of $50\%$ or more in the bolometric luminosity averaged over one to ten days.
We found these increased luminosities could be attributed to both unmeasured nuclei that feed into known nuclei (such was the case for the population of, for example, \iso{Ra}{224}), as well as unmeasured nuclei directly responsible for heating (as was the case, for example, for several isotopes of Rf). 

We found the behavior of the single trajectory calculations was reflected in the combined trajectories using the FRDM2012 mass model. 
In these cases, the competition between alpha decay and spontaneous fission with beta decay was not as large as in the single-trajectory cases, as all our combined trajectories had a substantial amount of material with electron fractions above 0.18. 
However, even in these circumstances, the description of the heating past one day is still incomplete without the contribution from fission and alpha decay.
Furthermore, the point in time at which alpha decay and fission begin to influence the overall magnitude of the heating differs, with NES contributions becoming relevant prior to one day, MLR at approximately one day, and MKT predicting a significant contribution closer to ten days.

Finally, we used our calculated abundances of the longest-lived isotopes of europium, thorium, and uranium to perform cosmochronometry calculations for a sample of six \rp-enhanced metal-poor stars. 
We found a larger uncertainty when we used actinide to europium ratios, as opposed to uranium:thorium ratios.
This is to be expected given the larger separation between europium and the actinides in the nuclear chart.
However, despite the large uncertainty, we were able to draw interesting conclusions.

One is that the use of different beta decay rates predicted disparities in the age estimated even from the actinide abundances alone. 
While there was significant overlap between the predictions resulting from NES and MLR actinide abundances, these differed from the MKT abundances, hinting at the extent to which these different beta decay rates hinder or facilitate actinide production.
Secondly, we were able to use the lanthanide abundances, together with the observational uncertainties to place a constraint on the age predictions, in the context of our model.
We showed that chronometric agreement depends on the beta decay rates.

We look forward to additional experimental efforts to measure beta decay properties \citep{gade_2016,aprahamian_2018,Tain_2018,Horowitz_2019,Savard_2020,Allmond_2020,Wu_2020,Schatz_2022}, which will greatly help to reduce this source of uncertainty in the predictions of kilonova light curves and of abundance predictions.  We also look forward to new theoretical predictions of  the thermodynamic conditions in merging neutron stars, of fission yields and daughter products, and of neutron capture and alpha decay rates, all of which have an important role to play.

\section{Acknowledgements}
This work was partially supported by the Fission in r-Process Elements (FIRE) topical collaboration in nuclear theory, funded by the U.S. DOE, contract No. DE-AC5207NA27344.
This work was partially supported by the Office of Defense Nuclear Nonproliferation Research \& Development (DNN R\&D), National Nuclear Security Administration, US Department of Energy.
This work was also possible due to support by the U.S. DOE through Los Alamos National Laboratory, operated by Triad National Security, LLC, for the National Nuclear Security Administration of the U.S. DOE. 
K.L. acknowledges support from the Seaborg Institute for funding under LDRD project 20210527CR, as well as from the Center for Nonlinear Studies.
J.E. acknowledges support from the Nuclear Computational Low Energy Initiative (NUCLEI) SciDAC-4 project under U.S. Department of Energy Grant No.\ DE-SC0018223 and by the U.S.
We acknowledge support from the NSF (N3AS PFC) grant No. PHY-2020275, as well as from U.S. DOE contract Nos. DE-FG0202ER41216 and DE-FG0295ER40934.
This research was supported in part by the National Science Foundation under Grant No. PHY-1430152 (JINA Center for the Evolution of the Elements).
This paper is approved for unlimited release, assigned LA-UR 22-28160.

\bibliography{ref}

\end{document}